# Tuning the optoelectronic properties of wide bandgap perovskites: Data-driven insights from combinatorial synthesis and high-throughput experimentation

Alexander Wieczorek[1], Sergey Tsarev[2,3], Nathan Rodkey[1], Oleksandr Pshyk[1], Stefanie Frick[1], Maksym V. Kovalenko[2,3], Sebastian Siol[1,*]

[1]Laboratory for Surface Science and Coating Technologies, Empa–Swiss Federal Laboratories for Materials Science and Technology Ueberlandstrasse 129, Duebendorf, CH-8600 Switzerland.
[2]Laboratory for Thin Films and Photovoltaics, Empa–Swiss Federal Laboratories for Materials Science and Technology Ueberlandstrasse 129, Duebendorf, CH-8600 Switzerland.
[3]Laboratory of Inorganic Chemistry, Department of Chemistry and Applied Biosciences, ETH Zürich, CH-8093 Zurich, Switzerland.

*Corresponding Author: sebastian.siol@empa.ch

## Abstract

The discovery and optimization of wide-bandgap lead halide perovskites (LHPs) is hindered by solution-based workflows with limited scalability. Large compositional parameter spaces present an additional challenge for materials optimization. Here, we establish an integrated, combinatorial workflow based on sequential physical vapor deposition that enables independent tuning of cation (Cs/Pb) and anion (Br/Cl) compositions. Applying automated structural, compositional, and optical characterizations across >500 samples regions of interest are rapidly screened in the quaternary Cs–Pb–Br–Cl space. From the screening, we establish a practical Cs/Pb window of 1.05–1.20 for wide bandgap perovskites, within which elevated PL yields were observed. Through in-depth analysis of the data set, we uncover a high-energy optical transition as a robust determinant for high PL yields. By combining mechanistic insight into the compositional origins of high PL efficiency with a fully integrated, high-throughput screening framework, and by openly releasing the complete multimodal dataset, this work provides a broadly accessible benchmark to accelerate data-driven discovery of wide-bandgap perovskites.

## Introduction

Wide-bandgap semiconductors (≥2 eV) are key to a wide range of optoelectronic applications. For instance, the stabilization and controlled growth of wide-bandgap nitride semiconductors has catalyzed the widespread use of light-emitting diodes (LEDs).[1] Their organic counterparts can be more easily miniaturized and are hence widely used in displays with self-emissive pixels. Still, the stabilization of OLEDs with wide bandgaps as required for wide-gamut displays remains challenging.[2,3] For medical and safety applications, wide bandgaps combined with high atomic weights are desirable for X-ray detection.[4] While commercially used CdZnTe-based detectors allow for high spatial resolution and low detection limits, complex production processes are required, resulting in high costs.[5]

Fully inorganic lead halide perovskite (LHP) semiconductors based on the $CsPbX_3$ stoichiometry (X = I, Br, Cl) with bandgaps of 1.7 – 3.0 eV are highly attractive candidates for optoelectronic applications.[6] For quantum light sources in particular, their nanocrystalline form may offer great benefits in the future.[7] To date, fully inorganic devices based on polycrystalline films have demonstrated excellent performances in X-Ray detection[8] and photodetection,[9] with continuously growing operational stability for perovskite LEDs.[10] Compared to their hybrid organic-inorganic counterparts, widely employed for solar cell applications, they have received far less attention to date. Automated processing and characterization of hybrid LHPs provide unique opportunities to optimize their composition[11] and processing conditions.[12] These workflows typically employ solution-based processing,[13] often based on inherently non-scalable deposition techniques, such as drop-casting. As the solubility of LHP precursors in common organic solvents tends to decrease with the resulting increasing bandgap,[14] similar solution-based approaches for inorganic LHPs are less feasible. Consequently, vapor-based deposition methods, such as chemical vapor deposition,[15] pulsed-laser deposition,[16] or thermal evaporation,[17] are more commonly reported for inorganic LHPs.[18] These methods are promising for the commercialization of perovskite optoelectronics, due to being more established in industrial thin-film deposition compared to solution-based processes.[19] Combinatorial vapor-based screenings are perfectly suited to close the gap between synthetic feasibility, scalability, and experimental throughput,[20] but have yet been underexplored for LHPs.[21]

Therefore, to unlock the full potential of inorganic LHPs, we establish a comprehensive combinatorial workflow that enables high-throughput exploration of compositional spaces. With combinatorial vapor-based depositions and a wide range of automated characterization platforms, we first screen the $Cs_xPbBr_{2+x}$ compositional space. From complementary characterization techniques, we gain insights into the chemical states, optoelectronic properties, and structural changes across compositional changes. This allows us to identify new trends and insights that are not obvious from conventional serial experimentation alone. Building up on our previous report,[22] the workflow maintains all samples under inert-gas during the characterization, thereby preventing external aging factors and enabling highly reliable data sets. While short-range order is often overlooked in high-throughput studies, we investigate it across all samples through X-ray photoelectron spectroscopy (XPS) coupled with charge-transfer data analysis, allowing for a more in-depth understanding of the observed trends in the optoelectronic properties. Furthermore, our synthetic protocol enables the independent screening of cation and anion composition in LHPs, vastly extending the usability of our screening workflow. In addition to probing the effect of cation composition, we co-vary the anion concentration via Cl doping to explore further compositions in the $Cs_xPb(Br_{1-y}Cl_y)_{2+x}$ space.

Through careful analysis of the resulting data set, spanning >500 unique LHP samples, we identify a distinct high-energy optical feature, that consistently correlates with enhanced photoluminescence. This feature can be related to the band dispersion diagram of pseudo-cubic $CsPbBr_3$, with our XPS measurements confirming lowered short-range order for related samples. It thus emerges as a new descriptor for experimental LHP screenings, with our openly available data guiding the optimization of wide-bandgap LHPs and data-driven efforts in this field.

## Results & Discussion

To ensure a reliable starting point for our combinatorial synthesis, we optimized our vapor-phase deposition using $CsPbBr_3$, an established inorganic LHP. For vapor depositions, both sequential and co-evaporation approaches of the $PbBr_2$ and CsBr precursors have been reported. Sequential approaches are more promising for large-scale applications, as they inherently limit cross-contamination during deposition and lead to highly structured thin films. This high degree of structural order has been associated with favorable charge transport properties in vapor-deposited optoelectronic devices.[23] Recent work by Lee *et al.* suggests that initiating deposition with a CsBr layer, rather than a Pb-based seed, promotes highly crystalline and phase-pure growth of fully inorganic lead halide perovskites.[24] This was shown to result from increased interfacial roughness between the cesium- and lead-based layers, which promotes the diffusion-controlled reaction during subsequent annealing. Likewise, Škorjanc *et al.* recently reported similar results for hybrid LHPs grown from CsCl seed layers.[25] In both cases, thermal annealing is required as the final step to convert the precursors to the LHP thin-film. Consequently, our synthetic protocol starts with the deposition of CsBr, followed by equimolar amounts of $PbBr_2$ and a final thermal annealing step.

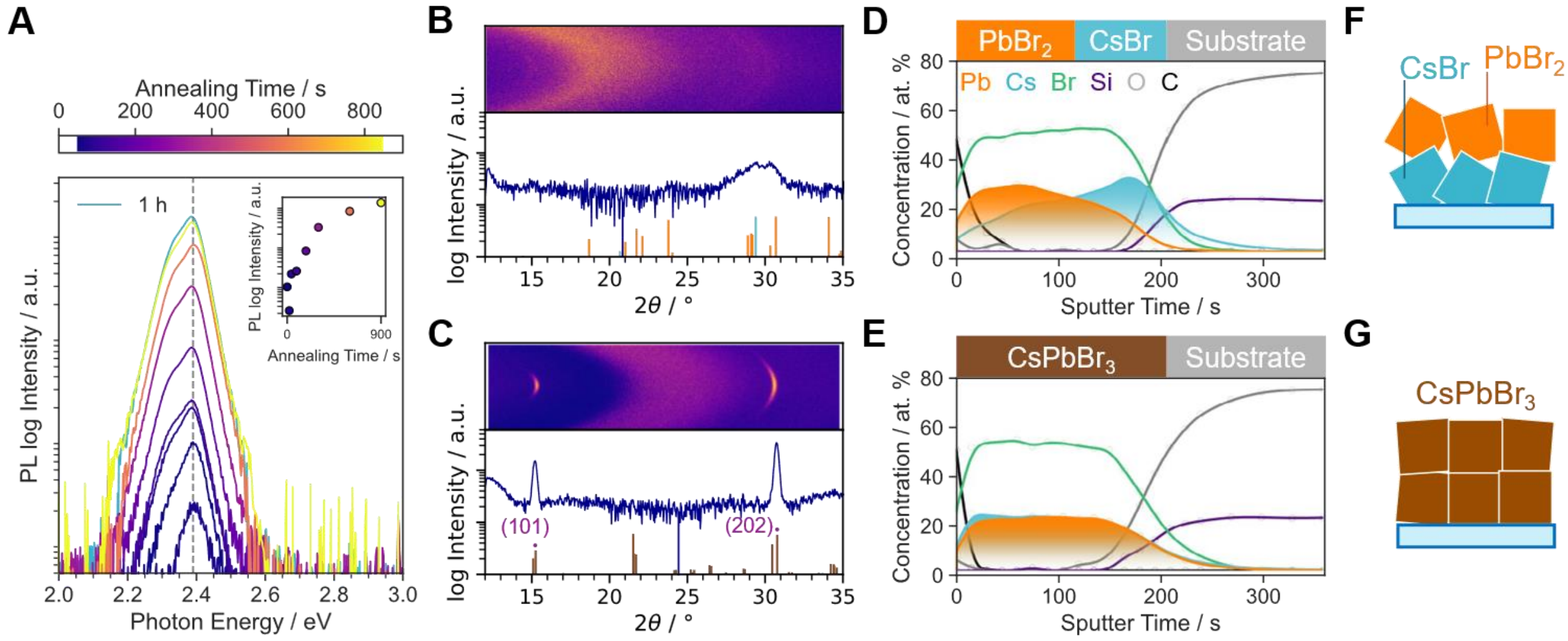

**Figure 1. A)** Annealing time optimization for sequentially evaporated $CsPbBr_3$ films. **B)** XRD diffractograms of as-deposited films and **C)** films annealed for 15 min at 250 °C. 2D diffractograms are shown on top, while the azimuthal-angle integrated 1D diffractogram is depicted below with reference peaks for $PbBr_2$ (orange), CsBr (lightblue), and $CsPbBr_3$ (brown). **D)** XPS depth profile for as-deposited films and **E)** films annealed for 15 min at 250 °C. **F)** Schematic illustration of the as-deposited and **G)** annealed thin-films.

As a figure of merit, we measured the resulting photoluminescence (PL) response of our films after varying annealing times as depicted in **Figure 1.A**. We observed a plateauing of the PL yield after 15 min of annealing at 250 °C with only negligible increases within the next 45 min. Longer annealed samples also resulted in a low-energy shoulder in the PL spectrum, commonly observed for LHPs, even in their single-crystalline state,[26] and not indicative of a second secondary phase, but typically related to photon reabsorption or cavity effects.[27,28]

Based on these optimized conditions, we characterized the structural properties of as-deposited and annealed films using 2D X-Ray diffractometry (XRD) measurements. Before annealing, as shown in **Figure 1.B**, we detected mixed phases of CsBr and $PbBr_2$ with low crystallinity as indicated by the overall peak signal intensity and peak broadness. This is in stark contrast to the annealed film, presented in **Figure 1.C**, which exhibited two distinct reflections related to the (101) and (202) planes of $CsPbBr_3$ (ICSD-14608). From the azimuthal angle intensity distribution, a strong increase in the preferential orientation was observed, suggesting the formation of highly textured thin-films after annealing.

To probe the compositional homogeneity along the film depth, we turned to X-ray photoelectron spectroscopy (XPS) depth profiling. Before annealing, as shown in **Figure 1.D**, the Pb content dominates at early sputter times and drops off as the Cs content rises towards the bottom of the film stack. Since the probing depth of XPS (≤10 nm) may broaden the measured interfacial separation, these results still suggest low intermixing between the CsBr and $PbBr_2$ phases. As shown in **Figure 1.E**, after annealing, we observed a near consistent ratio of the Cs, Pb, and Br content throughout the thin-film, implying the completed conversion to the perovskite phase. The absence of residual precursors even at the bottom of the film is noteworthy, as it is consistently observed for sequentially grown additive-free LHPs that are not fully vapor-deposited.[29,30]

As shown in **Figure 1.F** and **Figure 1.G**, these results suggest the presence of randomly oriented films with two layers after deposition that fully convert into a textured $CsPbBr_3$ film after annealing**.**

These annealing conditions yielded properties desirable for optoelectronic applications and were subsequently used for the synthesis of combinatorial material libraries. A full conversion of the films was indicated by the absence of any apparent secondary phases in the XRD mapping as well as an absence of substantial sub-gap absorption, which could also indicate scattering on secondary phases. Similarly, we do not see strong or systematic deviations from Vegardian lattice scaling which could indicate the formation of secondary phases. The PL intensity plateaus after the conversion and appears to be rather insensitive to non-optimized annealing conditions, so that changes in PL are expected to be governed by the materials composition, rather than variations in annealing conditions. While this is feasible and appropriate for the presented materials screening, further performance improvements may be realized by optimizing annealing protocols for specific compositions of interest in subsequent device development.

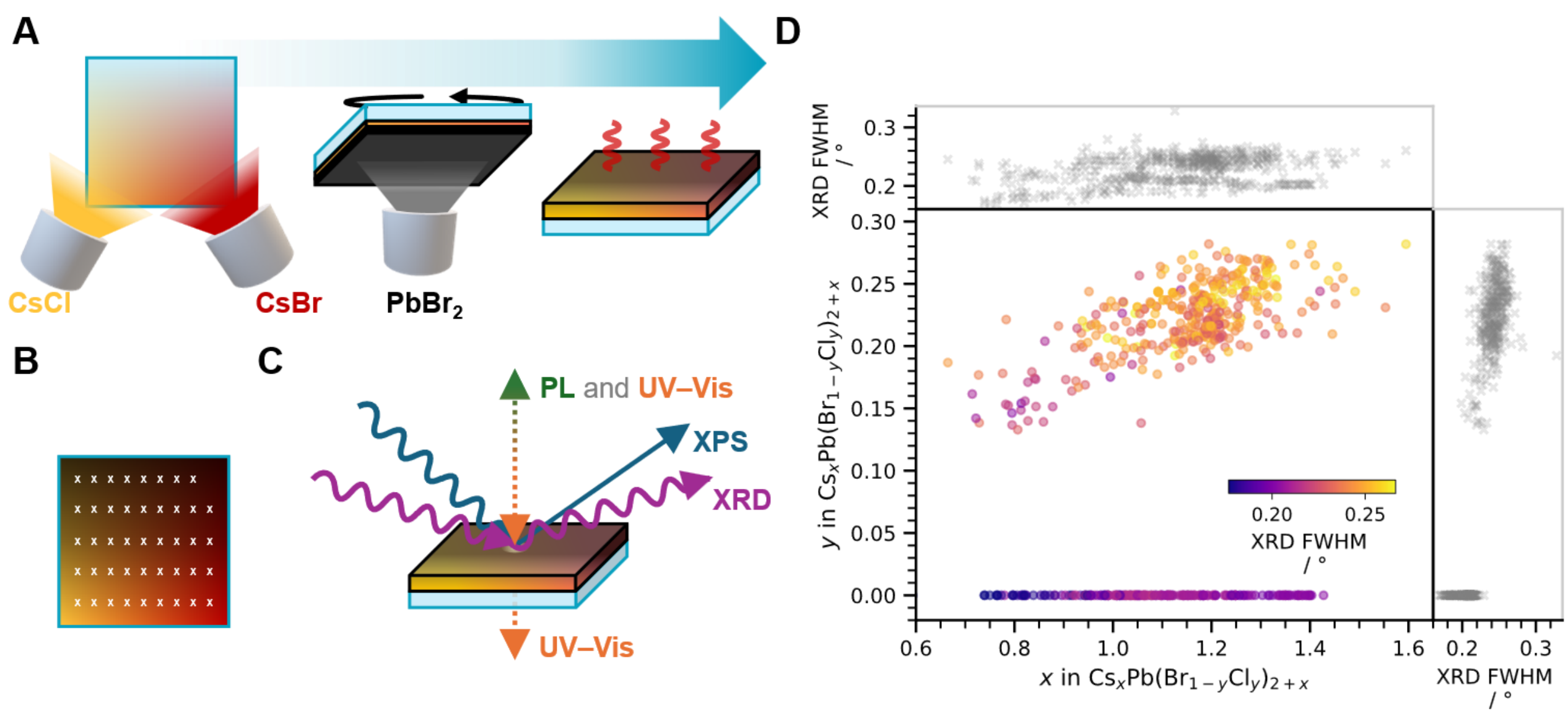

**Figure 2. A)** Combinatorial 2-step PVD deposition scheme used in this work: The first layer consists of a gradient deposition of either only CsBr or a co-deposition of CsBr, and CsCl. In the last deposition step, $PbBr_2$ is uniformly deposited across the substrate. After deposition, samples are treated with thermal annealing. **B)** Definition of samples on a Materials Library based on spatial coordinates. A blank sample without a thin film is placed on each Materials Library to simplify background measurements. **C)** Mapping-based methods used to characterize the Materials Library. **D)** Quaternary compositional map showing FWHM of the dominant XRD reflection between $2\theta \approx 30.7 – 31.1$ ° of all screened samples.

Through the extension of our synthetic approach, we unlocked the screening of the quaternary Cs-Pb-Br-Cl space as depicted in **Figure 2.A**: The process begins with a spatially controlled gradient deposition of CsBr, optionally complemented by CsCl introduced from a separate source within the evaporation chamber. This results in either a diagonal CsBr thickness gradient or a CsBr/CsCl composition gradient, overlaid with an orthogonal thickness gradient. Subsequently, a uniform $PbBr_2$ layer is deposited across the substrate, followed by thermal annealing, resembling the deposition and conversion procedure for $CsPbBr_3$ films shown in **Figure 1**. As pictured in **Figure 2.B,** this combinatorial method enables the fabrication of 44 distinct samples on a single substrate, further referred to as a Materials Library, under uniform processing conditions. With a custom sample holder, we can deposit up to 9 materials libraries simultaneously, increasing throughput to a maximum of 396 samples per PVD deposition (**Figure S1**).

To yield reliable insights into the resulting Materials Libraries with largely automated operation, we have adapted our existing high-throughput characterization workflow specifically for LHP thin-films: Each defined sample on the Materials Library is analyzed using XRD, XPS, UV–Vis, and PL on automated platforms, as summarized in **Figure 2.C**. Consequently, detailed insights into the compositional, structural, optical, and optoelectronic properties of each LHP sample are obtained. The experimentalist selects the ideal characterization platform for each investigation, enabling flexible high-throughput characterization for multiple application cases. Inert-gas conditions are fully maintained during this characterization workflow, ensuring the absence of degradation factors and thus more reliable data sets. Resulting experimental data from custom and commercial solutions is then standardized using a custom extract-load-transform pipeline. The resulting structured data set facilitates further data analysis in Python and enables open sharing of raw and analyzed data with the scientific community. As a result, the curated data set and complete code used for this investigation were published along this manuscript (see **Supplementary Information**).

Similar to the $CsPbBr_3$ films, we did not observe the formation of any crystalline secondary phases in the XRD patterns (**Figure S2**), indicating the full conversion of the precursors to the perovskite phase. Supplementary grazing-incidence X-ray diffraction measurements for a Cs-rich sample revealed a greater number of diffraction reflections, all of which could be attributed to the perovskite phase (**Figure S3**). A similar absence of secondary phases in XRD was previously reported by P. Becker *et al.* for CsI-rich $CsPbI_3$.[31] While both $CsPbBr_3$ and $CsPbCl_3$ share the orthorombic Pnma space group, the c lattice parameter changes significantly between $CsPbBr_3$ to $CsPbCl_3$ leading to a pronounced shift of the dominant reflection (**Figure S4**).[32] As depicted in **Figure 2.D**, from the full-width-half-max (FWHM) of this reflection, a broadening with increased Cl alloying in the $Cs_xPb(Br_{1-y}Cl_y)_{2+x}$ compositional space was observed, suggesting reduced crystallite sizes. This is contrary to the behavior of volatile organic chloride precursors, such as methylammonium chloride (MACl), which evaporate during thermal annealing. Among other effects, their addition mainly slows down the crystallization kinetics, which results in increased crystallite sizes.[33]

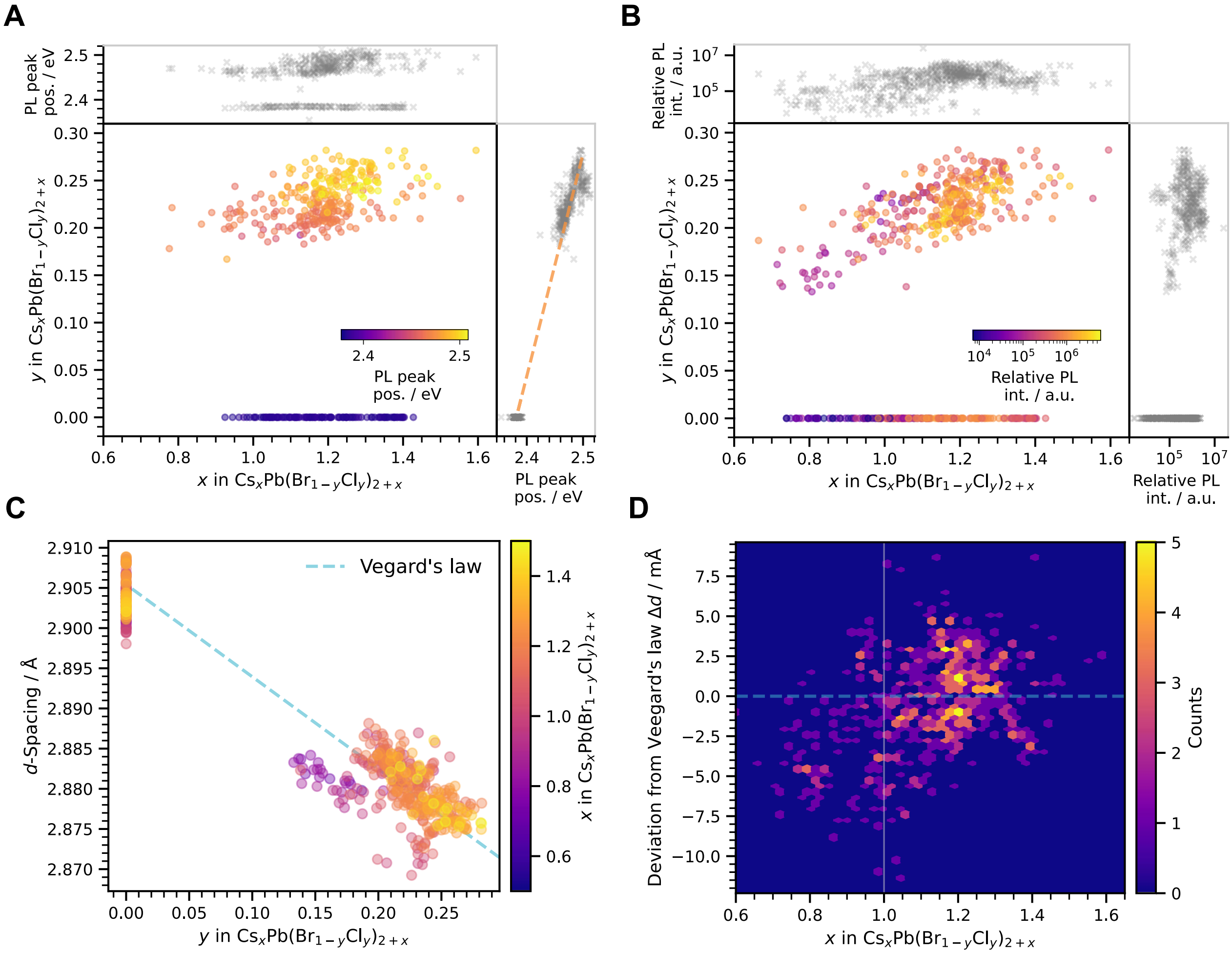

**Figure 3.** Influence of halide alloying on the anion site and Cs/Pb ratios for the cation site. **A)** Shift of PL peak positions with varied composition. **B)** Change of relative PL intensity with varying composition. **C)** Change of the (202) *d*-spacing in $CsPbBr_3$ with varying halide composition. The blue dashed line represents the expected *d*-spacing between the $CsPbBr_3$ and $CsPbCl_3$ endmembers, as predicted by Vegard's law. **D)** Deviation from Vegard's law against the Cs/Pb ratios across all halide compositions. The color scale indicates the number of individual measurements (Counts) for each bin, reflecting the local density of data points in this region.

We further confirmed the inclusion of Cl into the bulk perovskite lattice through the expected linear relationship between the Cl alloying fraction $y$ and the resulting PL peak position,[34] as shown in **Figure 3.A**. We note that understoichiometric (Cs/Pb < 1) samples with a Cs/Pb ratios of ≤ 0.9 exhibited exceptionally low PL responses and therefore prohibited the exact determination of the PL peak position. Through our combined UV–Vis and PL results, we calculated absorption-corrected PL peak area under curve values, to account for variations in absorption and film thickness, further denoted as relative PL intensity. Even though this metric is not an absolute measure of PLQY, it serves as a robust metric to compare the PL yields of different samples in this study. More details are given in the **Supplementary Information**. While the effect of Cl alloying on the relative PL intensity was minute (**Figure S5**), we observed variations of nearly two orders of magnitude across the screened Cs/Pb ratios, as depicted in **Figure 3.B**. The maximum of the relative PL intensity was within a Cs/Pb ratio range of 1.1 – 1.2, well into the overstoichiometric regime.

Strikingly, this ideal range for the A-site composition is maintained, regardless Cl rich alloying as apparent from Figure 3.B. To investigate whether this behavior correlates with structural changes in the material, we turned to a more detailed analysis of the X-ray diffractograms. As highlighted in **Figure 3.C**, we calculated the *d*-spacing for the most intensive XRD reflection within $2\theta$ angles of 30.7 – 31.1 ° for our samples. Using literature values of the end members, we compared these results to the expected *d*-spacing values for stoichiometric (Cs/Pb=1) halide alloys according to Vegard's law. The X-axis shows varying Cl alloying, whereas the Cs/Pb ratio is given as a color scale. The comparatively large spread around the ideal Vegardian lattice scaling (dashed blue line) can be explained by varying Cs/Pb ratios. For a given Cs/Pb ratio, the perovskite d-spacing closely follows Vegard’s law as a function of Cl alloying, indicating a predominantly single-phase solid solution. This would not be the case if significant amounts of secondary phases were present.[35] The small observed deviations from this trend may rather be explained by structural distortions (e.g. octahedral tilting), strain associated with defects or residual stress in the thin films.[36,37]

The calculated deviations from Vegard's law $\Delta d$ (i.e. the distance of the data points in Figure 3.C from the dashed blue line) are shown in **Figure 3.D**. These deviations could be linked to the Cs/Pb ratio within the thin films: For understoichiometric films, we observed mostly negative deviations from Vegard's law, in agreement with the increased stress within this regime. Within the Cs/Pb ratio regime of 1.0 – 1.3, we observed a flattening of this trend at $\Delta d \approx 0$ mÅ. This suggests relaxation of the crystal structure within this Cs/Pb regime, which encompasses the optimal regime for maximum relative PL intensities. At higher Cs/Pb ratios, we observed minute decreases of $\Delta d$, further resembling the previously discussed relative PL intensity trend.

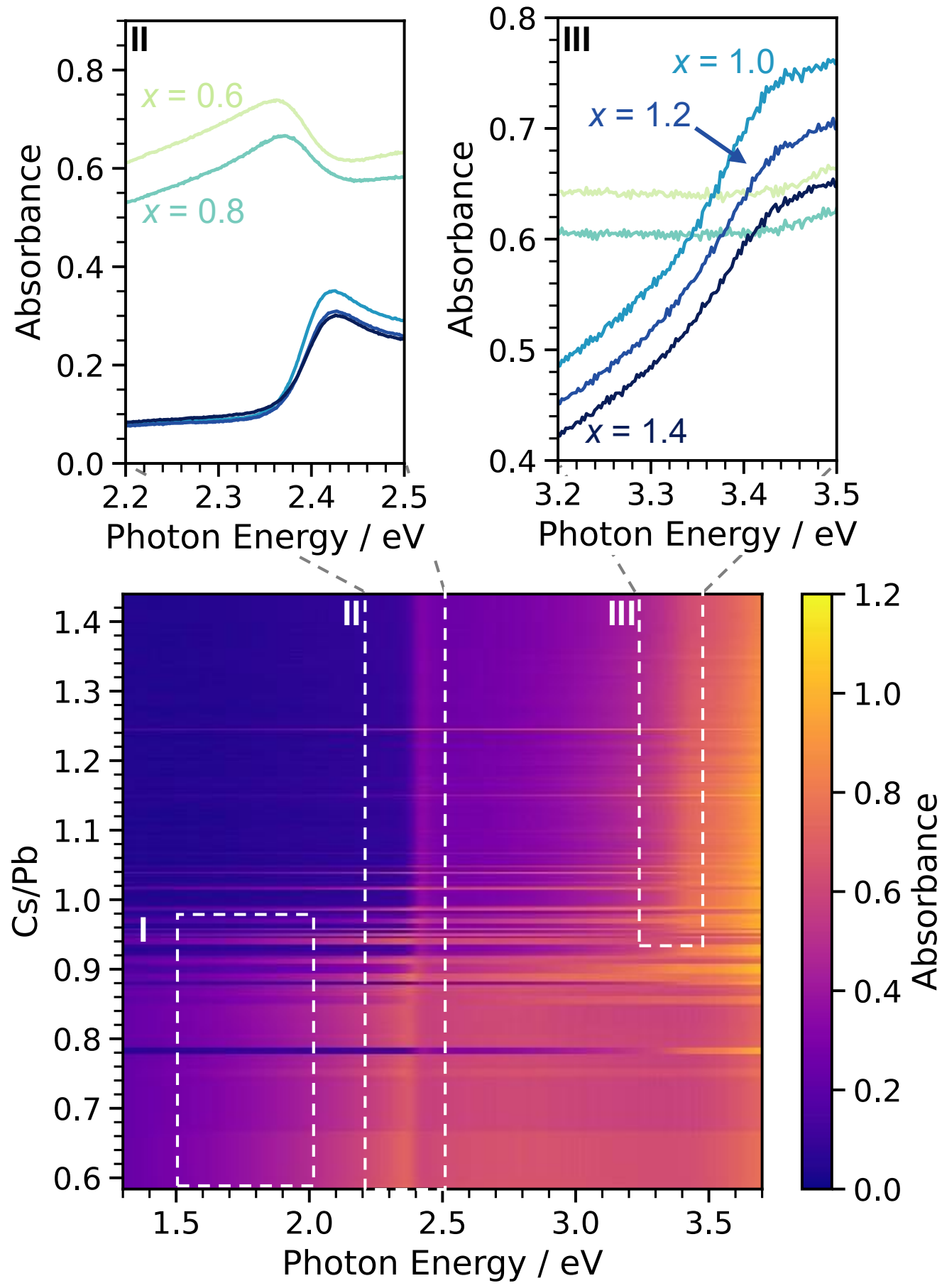

**Figure 4.** UV–Vis characterization of the $Cs_xPbBr_{2+x}$ compositional space. Three major spectral regimes were identified: **I)** Sub-bandgap absorption, **II)** Absorption onset, and **III)** UV-transition.

By performing UV–Vis measurements, we sought insights into changes of potentially non-emissive features and contributions of non-crystalline phases across the investigated compositional space. To simplify the analysis, we confined our spectral analysis to samples with Cs/Pb variations only, and hence within the $Cs_xPbBr_{2+x}$ compositional space.

As shown in **Figure 4**, an increased pseudo-absorbance within photon energies of 1.5 – 2.3 eV (**regime I**) was observed for understochiometric $Cs_xPbBr_{2+x}$ films (**Figure S6**). This typically results from increased light scattering, often caused by precipitation or delamination within the films.[38,39] Indeed, from our XPS results, we found profound Si 2p core level features related to the underlying substrate within this compositional window (**Figure S7**). Within the 1≤Cs/Pb≤1.5 compositional window, we observed a complete lack of this feature, suggesting the formation of closed homogeneous thin-films.[40]

For films that are not understoichiometric, we observed a sharp absorption onset at photon energies of approximately 2.4 eV (**regime II**). The prominent absorption peak was previously assigned to discrete excitonic state levels below the intrinsic band gap.[41] We determined the onset of this transition through its inflection point, a method that we recently reported to be robust against sub-bandgap absorption and scattering in high-throughput workflows.[22] Furthermore, it provides a more robust determination than the conventional Tauc plot method for automated data processing, as the latter can introduce large uncertainties on the order of multiple 0.1 eV.[42] Across all measured samples, we observe a nearly constant shift between the absorption onset and the PL maximum (**Figure S8**), suggesting a similar Stokes shift within the screened compositional window.

Beyond this feature, an additional feature, energetically situated at approximately 3.4 eV, was detected (**regime III**), which was predominantly found for slightly overstoichiometric films. This transition, hereafter referred to as the *UV-transition*, has received limited attention in polycrystalline $CsPbBr_3$.

For the full UV–Vis data set of the $Cs_xPb(Br_{1-y}Cl_y)_{2+x}$ compositional space, we observed the same three regimes, albeit energetically shifted with respect to the halide content present in the films (**Figure S9**).

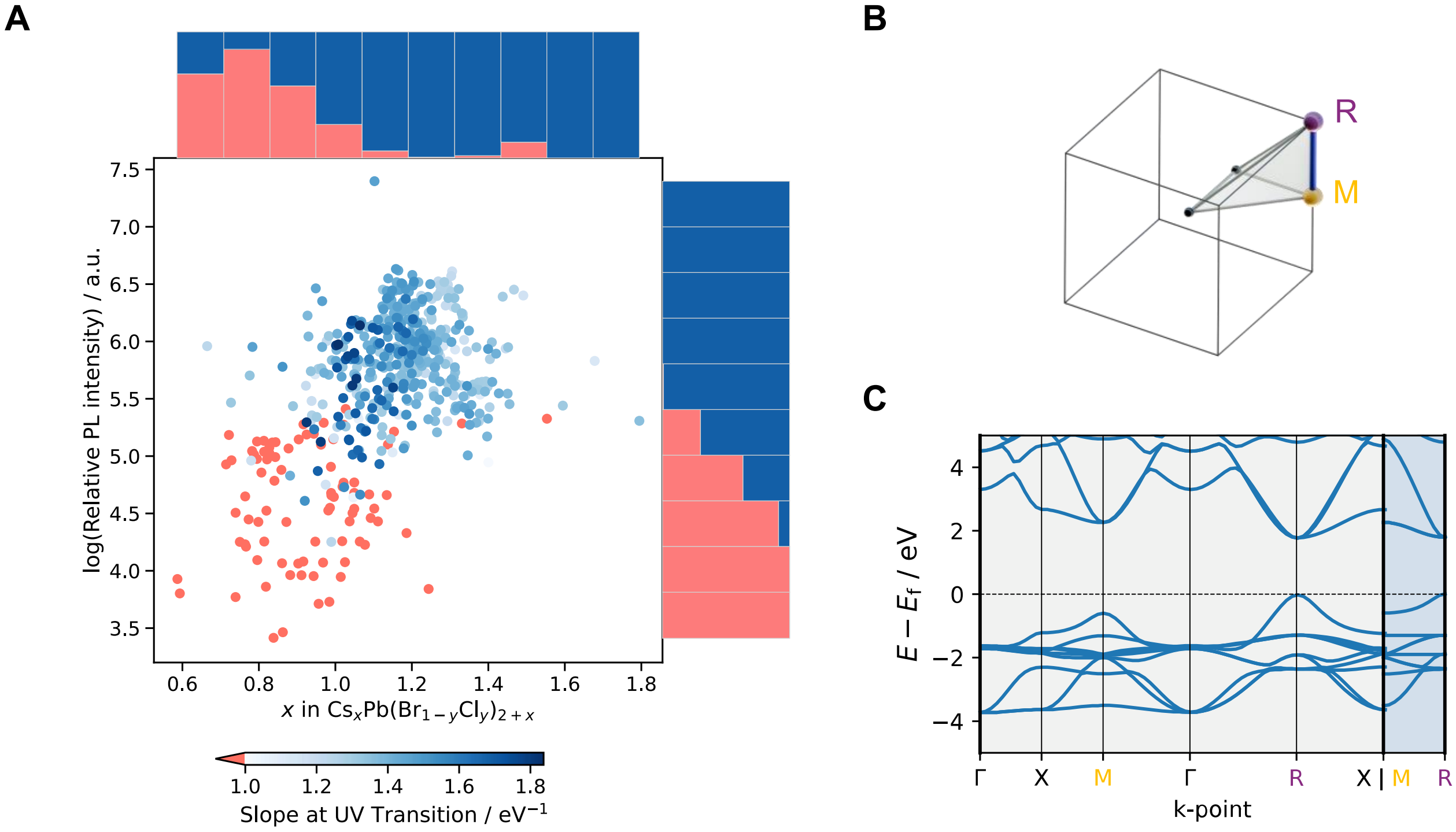

**Figure 5. A)** Relative PL intensities across all screened samples with varying Cs/Pb ratios. The significant presence of a UV-Transition is further quantified by the magnitude of its absorbance slope. Red colored points mark samples exhibiting a UV-transition slope of < $1 eV^{-1}$. **B)** Scheme of the first Brillouin zone of cubic $CsPbBr_3$ with the M-, and R-points highlighted in yellow, and violet, respectively. **C)** Band dispersion diagram of cubic $CsPbBr_3$ extracted from Materials Project.[43] The blue highlighted region marks the k path between the M- and R-points.

Strikingly, as highlighted in **Figure 5.A**, the previously introduced UV-transition emerges as a key determinant of the resulting photoluminescence intensities. A clear correlation between the relative PL intensity and the Slope at the UV transition can be observed. Notably, we observe a partial separation between samples with and without a UV-transition around a Cs/Pb ratio of 1, matching with the largest variation in observed relative PL intensities. This spread can be attributed to the presence or absence of a UV transition, highlighting its significant correlation with the optical emission characteristics. Our combinatorial synthesis equalizes deposition conditions (e.g., substrate temperature or precursor impurities), but differences in microstructure and thickness across the materials libraries may contribute to some of the observed spread. However, these residual strong variations could explain the often inconsistent reports on LHP thin-film properties, underscoring reproducibility challenges within the field.[44]

Multiple mechanisms explaining an optical absorption feature at 3.4 eV have been suggested before for the Cs-Pb-Br compositional space. For instance, in wet chemical synthesis this optical transition was previously linked to solvated $[PbBr_4]^{2-}$ complexes after their successful isolation with surface ligands.[45] These compounds act as a precursor to $Cs_4PbBr_6$, a potential side product during $CsPbBr_3$ nanocrystal synthesis, which favorably forms under Cs-rich conditions (**Figure S10**).[46] However, $Cs_4PbBr_6$ impurities exhibit an additional higher energy absorption onset at 3.9 eV, a transition which was absent for most samples exhibiting the UV-transition in the present work (**Figure S11**). At high

Cs-ratios with $x$>1.2, we observed the gradual decrease of the UV-transition and the $Cs_4PbBr_6$-related absorption onset, suggesting a different mechanism underlying this feature in our investigation.

Chen *et. al* reported the clear prominence of the UV-transition in single-crystalline $CsPbBr_3$,[47] suggesting that it results from a mechanism distinct from impurity-related effects. In addition, the UV transition is commonly observed even in the UV–Vis spectra of phase-pure $CsPbBr_3$ nanocrystals.[6,48] Hence, the presence of this optical absorption feature does not necessarily suggest the formation of additional phases.

Indeed, our XRD screening suggests the relaxation of the crystal structure within this compositional regime and thus reflected structural properties of single crystals and nanocrystals. Guided by the detailed optical investigation of a $CsPbBr_3$ single crystal,[47] we therefore assign this optical feature to an optical transition at the M-point in the Brillouin zone for pseudo-cubic $CsPbBr_3$, as presented in **Figure 5.B**. Besides the intrinsic band gap at the R-point, it marks the second-lowest optical transition in the band dispersion diagram shown in **Figure 5.C**, accounting for the observed strong optical absorption.

We note that multiple direct optical transitions can be found in the band dispersion diagram of orthorhombic $CsPbBr_3$ as well, which are linked to similar energies (**Figure S12**). Indeed, G. Mannino *et al.* recently investigated the phase transition of $CsPbBr_3$ single crystals from orthorhombic to tetragonal by ellipsometry.[49] In all cases, the presence of an optical transition at 3.4 eV was observed with minor energetic shifts. Although no explanation was given for the cause of this feature, the fact that the UV transition was observed also in single-crystal materials underlines that its origin is intrinsic in nature. Crucially, its amplitude was determined to approximately double with the phase transition from orthorhombic to cubic.

Combining these prior literature results with our investigation, we hypothesize its strength follows increased symmetry within the crystal lattice and hence lowered structural disorder. For LHPs, this may be linked to the octahedral tilting within the perovskite structure. A recent investigation by Chiang *et al.* on high-entropy alloy nanocrystals derived from $CsPbBr_3$[50] showed a similar loss of UV–Vis features with compositional disorder induced by B-site alloying.

While all measured XRD diffractograms suggest long-range orthorhombic order, the short-range order may still appear pseudo-cubic. For high-throughput combinatorial characterizations of this short-range order, our group recently demonstrated the high potential of XPS analysis on amorphous ceramics.[39] Similar to the UV–Vis analysis, we commenced our analysis with the analysis of the $Cs_xPbBr_{2+x}$ compositional space to isolate the effect of Cs/Pb ratio changes alone.

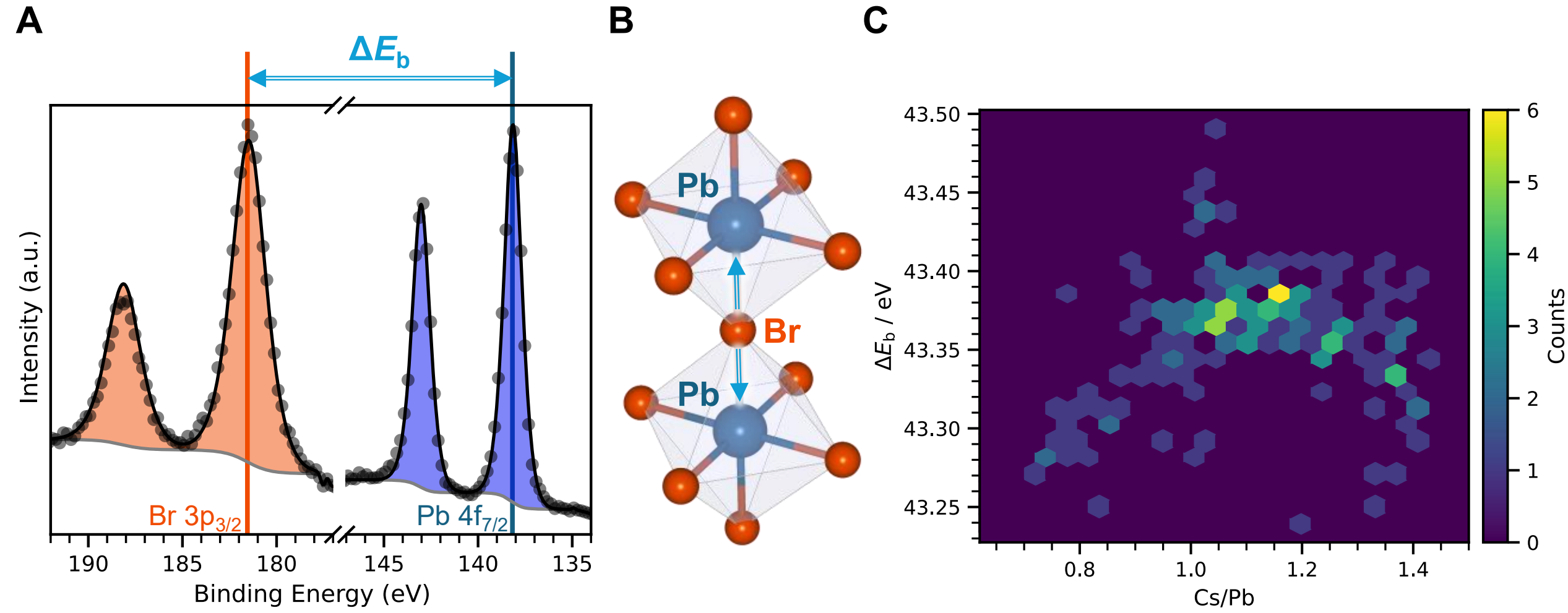

Figure 6. A) Br 3p and Pb 4f core level regions with extraction of relative binding energy difference $\Delta E_b$ for charge transfer analysis. B) Scheme of relevant nearest neighbor interactions for charge transfer analysis between Pb and Br. The electron density at the Pb-site is dependent on the Pb-Br-Pb angle. C) Changes of $\Delta E_b$ over varied Cs/Pb ratios across the $Cs_xPbBr_{2+x}$ compositional space. The color scale indicates the number of individual measurements (counts) for each bin, reflecting the local density of data points in this space.

Consequently, we performed chemical state analysis based on the Cs 3d, Pb 4f, and Br 3p core level regions. It is noteworthy that, in addition to structural modifications of the material, changes in surface chemistry might affect the PL intensity, for instance when they result in different degrees of surface passivation. While such an effect cannot be completely ruled out, the XPS analysis of the samples revealed no signs of secondary phases on the surface. For all elements, we observed a single dominant core-level feature and therefore chemical state in both under- (**Figure S13**) and overstochiometric (**Figure S14**) films. None of the investigated samples exhibited a Fermi-edge in the valence band region (**Figure S15**), indicating the absence of metallic fractions (*e.g.*, $Pb^0$) as sometimes observed for degraded LHP films.[51]

While minor absolute binding energy shifts are often interpreted, charging effects may induce non-chemical shifts for semiconductors, especially.[52,53] The widely employed C 1s referencing still results in errors of the binding energy scale, due to insufficient electrical contact of the adventitious carbon and the underlying sample surface. These errors, typically reported to be in the range of ±0.25 eV[54] but reaching up to 2 eV,[55] hinder the usage of absolute binding energies in XPS for robust chemical state analysis for LHPs in particular.

Consequently, analysis based on relative binding energy shifts has emerged as a more reliable approach for the chemical state analysis of semiconductors.[53] For Pb-based metal halide perovskites, we have recently introduced this concept based on the charge-transfer analysis of the Pb- and X-sites.[56] As shown in **Figure 6.A**, the relative binding energy $E_b$ of the Pb $4f_{7/2}$ core level maximum is compared that of its constituent halide-site, i.e., Br $3p_{3/2}$, resulting in a relative binding energy value $\Delta E_b$:

$$\Delta E_\mathrm{b} = E_\mathrm{b}[\mathrm{Br\ 3p_{3/2}}] - E_\mathrm{b}[\mathrm{Pb\ 4f_{7/2}}] \quad (1)$$

Due to the binding energy range for both core levels, a shift towards higher values indicates an increased electron density at the Pb-site compared to the Br-site. This electron density and hence shift is highly dependent on the nearest neighbor interactions. For LHPs, these are largely governed by the tilting of the $[PbBr_6]^{4-}$ octahedra towards each other, as visualized in **Figure 6.B**.

Performing this analysis on our combinatorial data set yields the results depicted in **Figure 6.C**, with a peak of $\Delta E_b \approx 43.4$ eV at a Cs/Pb ratio of approximately 1.1. Notably, this value lies within 70 meV of the value for $CsPbBr_3$ nanocrystals synthesized based on a ligand-assisted reprecipitation protocol in a recent study by Y.T. Chiang *et al.* and our group.[50] This resemblance suggests that Cs-rich films share similar nearest neighbor interactions and electron density at the Pb-site, indicative of reduced octahedral tilting. Collectively, these results indicate a higher structural order at a Cs/Pb ratio range of 1.05 – 1.2.

Samples containing Cl similarly only exhibit a single dominant chemical state in their core-level regions (**Figure S16**). By restricting the analysis to the previously determined optimal Cs/Pb ratio range, we probe the secondary effect of the Cl inclusion on the relative binding energy. As expected for increased bond ionicity in the $[PbX_6]^{2-}$ octahedra, Cl-alloyed samples exhibit a minute decrease in $\Delta E_b$, reflecting a marginally lower electron density at the Pb site (**Figure S17**). We note, however, that these effects are on the order of $10^{-2}$ eV and are thus minute compared to the effects induced by changes in the Cs/Pb ratio in this investigation.

Collectively, these findings mirror the Cs/Pb range of 1.05 – 1.2 as optimal for maximizing photoluminescence in LHP films. Our combinatorial XPS screening solidifies our hypothesis that these effects can be attributed to the minimum in short-range disorder within this compositional window.

## Conclusion

In this work, we have developed a fully integrated, high-throughput workflow that seamlessly couples scalable Cs–Pb–Br–Cl thin-film deposition with automated structural, compositional, and optical screening across several materials libraries containing a total of 569 samples. Through in-depth analysis of this multimodal data set, we identified the UV-transition feature as a reliable predictor of strong photoluminescence, linking it to the M-point transition in the band dispersion diagram of pseudo-cubic $CsPbBr_3$. XPS-based charge transfer analysis indicates reduced octahedral tilting in the compositions exhibiting this transition, suggesting minimized short-range structural disorder as the underlying origin. Importantly, this UV-transition signature remains consistent across different halide mixtures, making it a broadly applicable proxy for optimizing wide-bandgap LHPs that can be employed in the design of light-emitting devices.

The high-throughput screening workflow presented in this work can be augmented with other tools and provides a solid foundation for future device-screening studies that explore the feasibility of the investigated materials.

By releasing the complete, multimodal data set, we provide both a benchmark for reproducible LHP screening and a transferable resource for data-driven materials discovery and optimization. Because the UV-transition can be rapidly detected with standard optical tools, this work enables universally accessible, mechanism-guided experimental screening strategies across perovskite chemistries. We further anticipate that the concepts and workflow presented here will extend to other semiconductor families where short-range disorder affects optoelectronic properties.

# Methods

## Data Analysis

To structure and reference the data sets acquired from the different characterization platforms, a customized version of COMBIgor was used.[57] Following this extraction, load, transform (ELT) pipeline, the COMBIgor projects were saved as HDF5 files. Advanced data analysis was then performed using custom modules developed within a Python 3.12 environment.

## Synthesis of Materials Libraries

The materials libraries were manufactured using thermal evaporation of CsBr, CsCl, and $PbBr_2$ precursors. Generally, Cs-based precursors were deposited first, followed by sequential deposition of $PbBr_2$. The evaporation rates were controlled through quart-crystal microbalances calibrated for target thicknesses under substrate rotation.

For screening the Cs–Pb–Br compositional space, CsBr was first evaporated without substrate rotation, followed by equimolar deposition of $PbBr_2$ with rotation enabled. To explore the Cs–Pb–Br–Cl compositional space, the first step was modified to a co-evaporation of CsBr and CsCl. The evaporation rates of the Cs-based precursors were adjusted to yield equimolar deposition across the substrate. Since the precise compositional gradients are hard to predict prior to the deposition the composition of every sample on each materials library was measured via inert-gas XPS experiments.

## Characterization of combinatorial libraries

Each combinatorial materials library contains 44 individual samples, which are defined by a coordinate grid as reported in our prior work.[22] Each coordinate on the library corresponds to a sample with a unique composition. All measurements are carried out in an automated fashion for each coordinate (i.e. sample).

### Optical characterizations and coupled *in-situ* analysis

Automated UV–Vis and photoluminescence (PL) spectra of the pristine libraries were acquired using a self-built setup. While the specifics of the UV–Vis characterization optics have been reported elsewhere,[22] the setup was extended with a secondary laser optics for PL measurements. For the excitation beam, a 405 nm single-mode fiber-coupled diode laser was used. The fiber was terminated with an aspheric collimator and the laser beam routed through a 405 nm bandpass filter. The laser beam was then reflected using a fused silica dielectric mirror as well as a longpass dichroic mirror and focused on the materials library using an achromatic doublet lens. Emission light was further focused from the same lens, transmitted through the longpass dichroic mirror and finally coupled into a multimode fiber using a high $N_A$, achromatic fiber collimator. A charge-coupled device (CCD) detector was used to rapidly record the resulting PL spectra. To block stray light, the entire measurement setup is enclosed in a box made from black anodized aluminum. The excitation wavelength, collection geometry, and integration time are kept constant across all samples. To account for variations in film thickness and absorption of samples with varying composition we define an absorption-corrected relative PL intensity (see Supporting Information) to allow for a robust comparison of the PL yield for different samples in our data set.

**X-Ray diffraction (XRD)**

X-Ray diffraction (XRD) was performed in a custom-made inert-gas dome mounted in a Bruker D8 Discovery, equipped with a Cu Kα radiation source and within a Bragg-Brentano geometry. Automated measurements were performed by mapping the complete materials library on the xy-stage of the diffractometer. Reference diffractograms were calculated from the respective .cif files acquired from the Inorganic Crystal Structure Database (ICSD) release 2024.2 using the Python Materials Genomics (pymatgen) package[58] within a Python 3.12 environment.

**X-ray photoelectron spectroscopy (XPS)**

X-ray photoelectron spectroscopy (XPS) was performed in a PHI Quantum system to which samples on the insulated substrates used for the materials libraries were transferred using a custom-made inert-gas transfer vessel in Ar atmosphere. Automated measurements were performed by mapping each one spot per sample on the materials library. XPS measurements were performed at a pressure of $10^{-9}$ − $10^{-8}$ Torr. The monochromatic Al Kα radiation was generated from an electron beam at a power of 24.7 W and a voltage of 15 kV. Charge neutralization was performed using both low-energy electron and $Ar^+$ beam sources. Short-term measurements (<1 min) of the Pb 4f core level before and after each presented long-term measurement (~30 min) were conducted to rule out changes in the chemical state due to potential X-ray induced beam damage. To determine the composition of the samples, batch peak fitting of photoelectron features was performed in Casa XPS using Voigt profiles with Gauss-Lorenz (GL) ratios of 60 following Shirley-background subtraction. The binding energy scale was referenced to the main component of adventitious carbon at 284.8 eV, typically resulting in an inaccuracy of approx. ±0.25 eV.[54] The results from our charge transfer analysis are not affected by this inaccuracy. Atomic ratios were calculated using the instrument specific relative sensitivity factors (RSF).

Depth profiles were acquired by etching the samples with an ion beam of $Ar^+$ ions accelerated with 2 kV and rastered across a $2 \cdot 2$ $mm^2$ area between each XPS measurement cycle.

## Acknowledgements

A.W. and S.S. acknowledge funding from the Strategic Focus Area–Advanced Manufacturing (SFA–AM) through the project Advancing manufacturability of hybrid organic–inorganic semiconductors for large area optoelectronics (AMYS). A.W. and S.S. further acknowledge financial support by the ETH ORD program for the development of data management tools. S.S. and N.R. acknowledge the Swiss National Science Foundation under Projects 226588 and 10004403. O. P. and S. S. acknowledge funding from the SNSF under project 227945. S. T. and M. K. support by ETH Zürich through the ETH+ Project SynMatLab: Laboratory for Multiscale Materials Synthesis. A.W. acknowledges Sasa Vranjkovic and the Empa tool shop for their help in the technical construction and manufacturing of CNC-machined parts required for this work. A.W. further acknowledges Dr. Simon Christian Böhme for helpful discussions and introductions to optical characterizations.

## Conflict of Interest Statement

The authors declare no conflicts of interest regarding this manuscript.

## Data availability statement

The data and code needed to reproduce the findings discussed in this article are available on https://github.com/Empa-CT/Open_WBG_Perovskite_Map. *(The repository will be published after acceptance of the article.)*

*Supporting Information for*

# Tuning the optoelectronic properties of wide bandgap perovskites: Data-driven insights from combinatorial synthesis and high-throughput experimentation

Alexander Wieczorek[1], Sergey Tsarev[2,3], Nathan Rodkey[1], Oleksandr Pshyk[1], Stefanie Frick[1], Maksym Kovalenko[2,3], Sebastian Siol[1,*]

[1]Laboratory for Surface Science and Coating Technologies, Empa–Swiss Federal Laboratories for Materials Science and Technology Ueberlandstrasse 129, Duebendorf, CH-8600 Switzerland.

[2]Laboratory for Thin Films and Photovoltaics, Empa–Swiss Federal Laboratories for Materials Science and Technology Ueberlandstrasse 129, Duebendorf, CH-8600 Switzerland.

[3]Laboratory of Inorganic Chemistry, Department of Chemistry and Applied Biosciences, ETH Zürich, CH-8093 Zurich, Switzerland.

*Corresponding Author: sebastian.siol@empa.ch

# Table of Contents

# List of Figures

## Determination of relative PL intensity

The photoluminescence quantum yield (PLQY) characterizes the efficiency of photoluminescence (PL) for a given material. It is mathematically defined as:

$$PLQY = \frac{n_{\text{emitted}}}{n_{\text{absorbed}}} \tag{S1}$$

, where $n_{\text{absorbed}}$ denotes the number of absorbed photons and $n_{\text{emitted}}$ denotes the number of emitted photons.

Through our combined UV–Vis and PL characterizations, we obtained a relative photoluminescence quantum yield $\text{PLQY}_{\text{rel.}}$ according to:

$$PLQY \propto \frac{E_{\text{PL,emission}}}{A_{\text{PL,excitation}}} \propto \frac{\int h\nu\, I_\nu\, d\nu}{A_{\text{PL,excitation}}} = PLQY_{\text{rel.}} \tag{S2}$$

, where $E_{\text{PL,emission}}$ denotes the energy of the emitted photons, $A_{\text{PL,excitation}}$ denotes the Absorption at the excitation wavelength from UV–Vis, and $I_{\text{V}}$ denotes the relative PL intensity as a function of photon frequency.

## Supplementary Figures

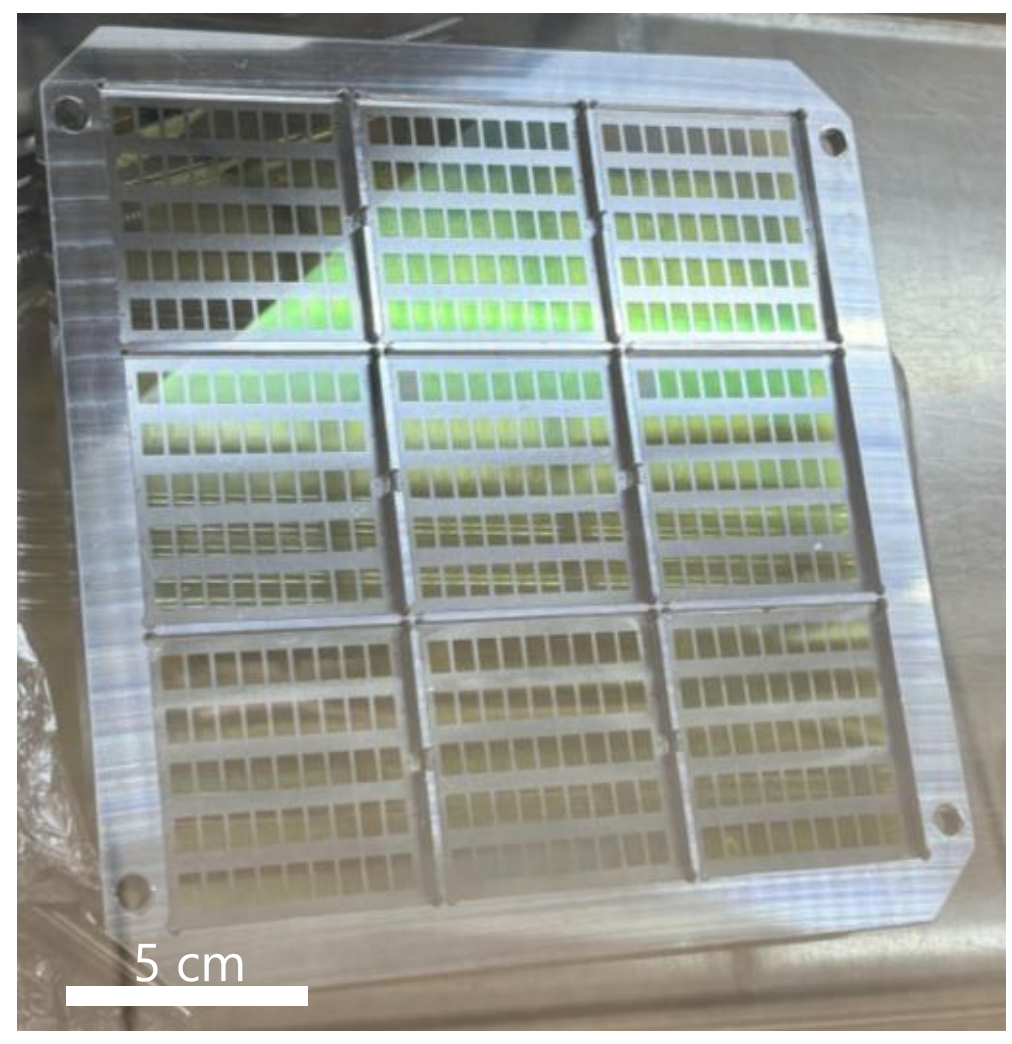

**Figure S 1.** Picture of a set of 9 Materials Libraries after deposition.

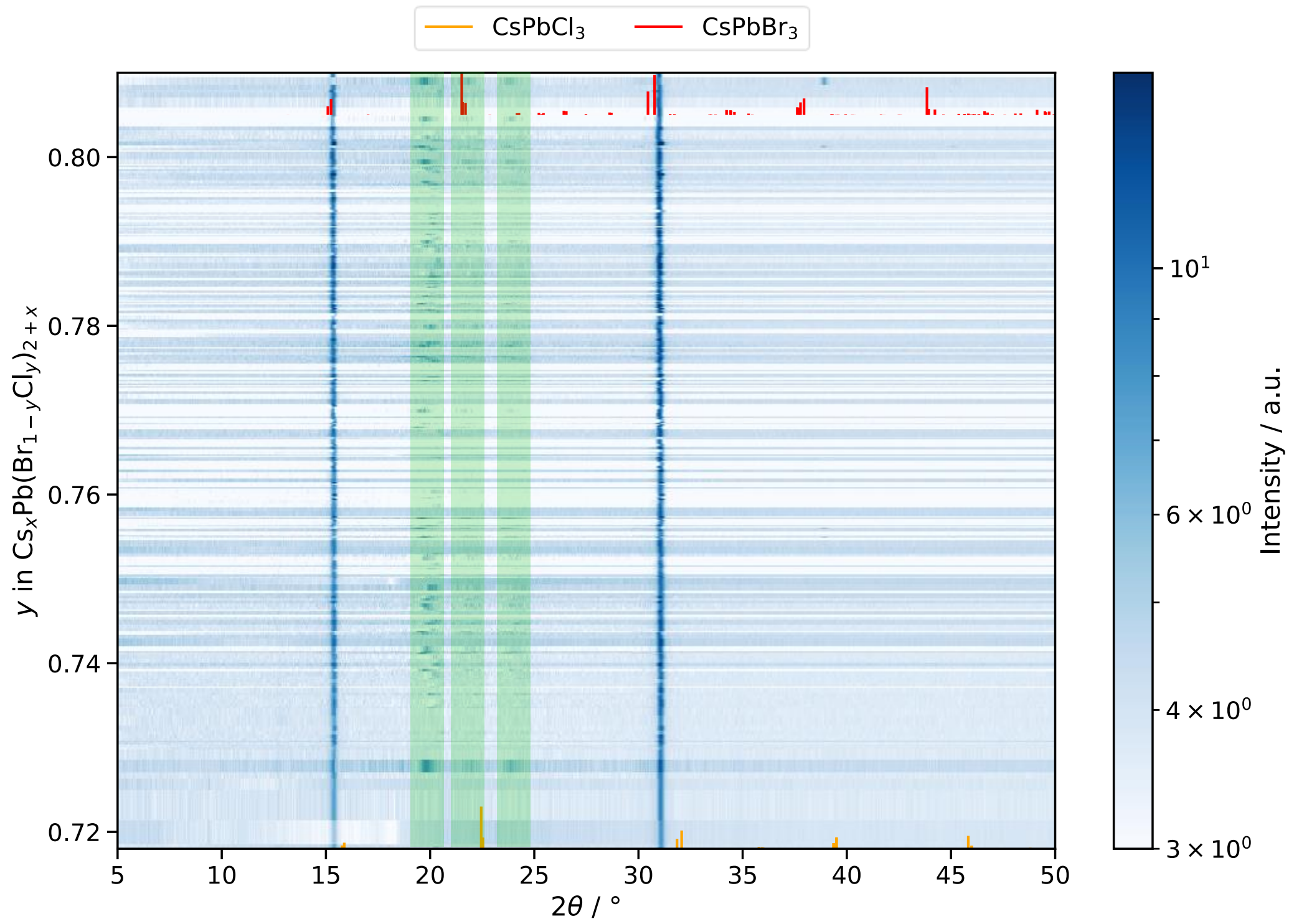

**Figure S2.** False-color XRD diffractogram for samples with varying halide ratios with references for $CsPbBr_3$ (ICSD-14608) and $CsPbCl_3$ (ICSD-143611). The green shaded area highlights regions where peaks from the used inert-gas XRD dome may be expected.

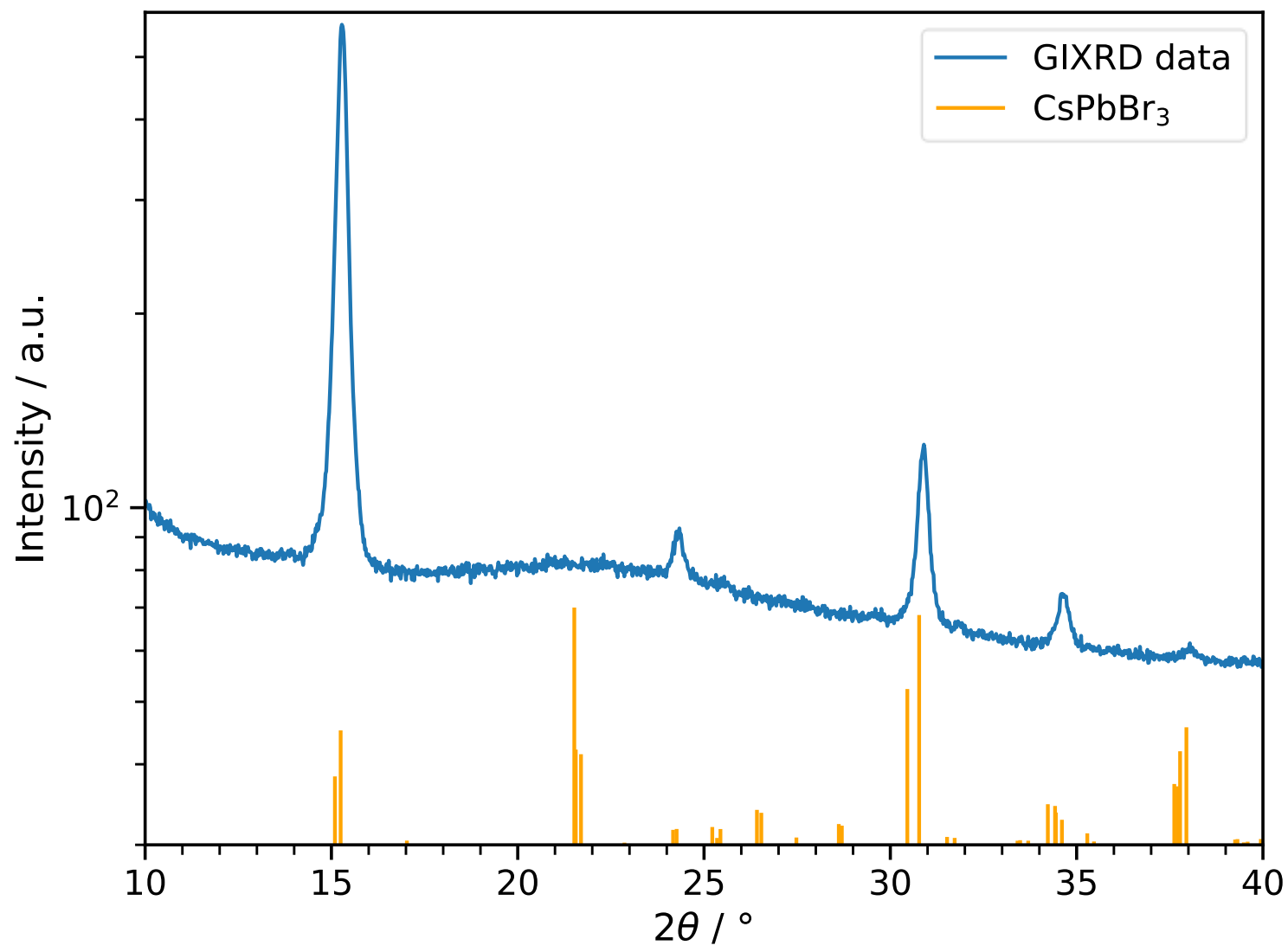

**Figure S3.** GIXRD diffractogram of a representative sample within the $Cs_xPbBr_{2+x}$ compositional space. Simulated powder diffraction patterns of $CsPbBr_3$ (ICSD-14608) are shown as vertical lines. In contrast to the XRD measurement, where only the (110) and (220) reflections were found, additional reflections, including the (210) and (004) reflections, were observed in GIXRD. Notably, no reflections indicating the presence of secondary phases were identified.

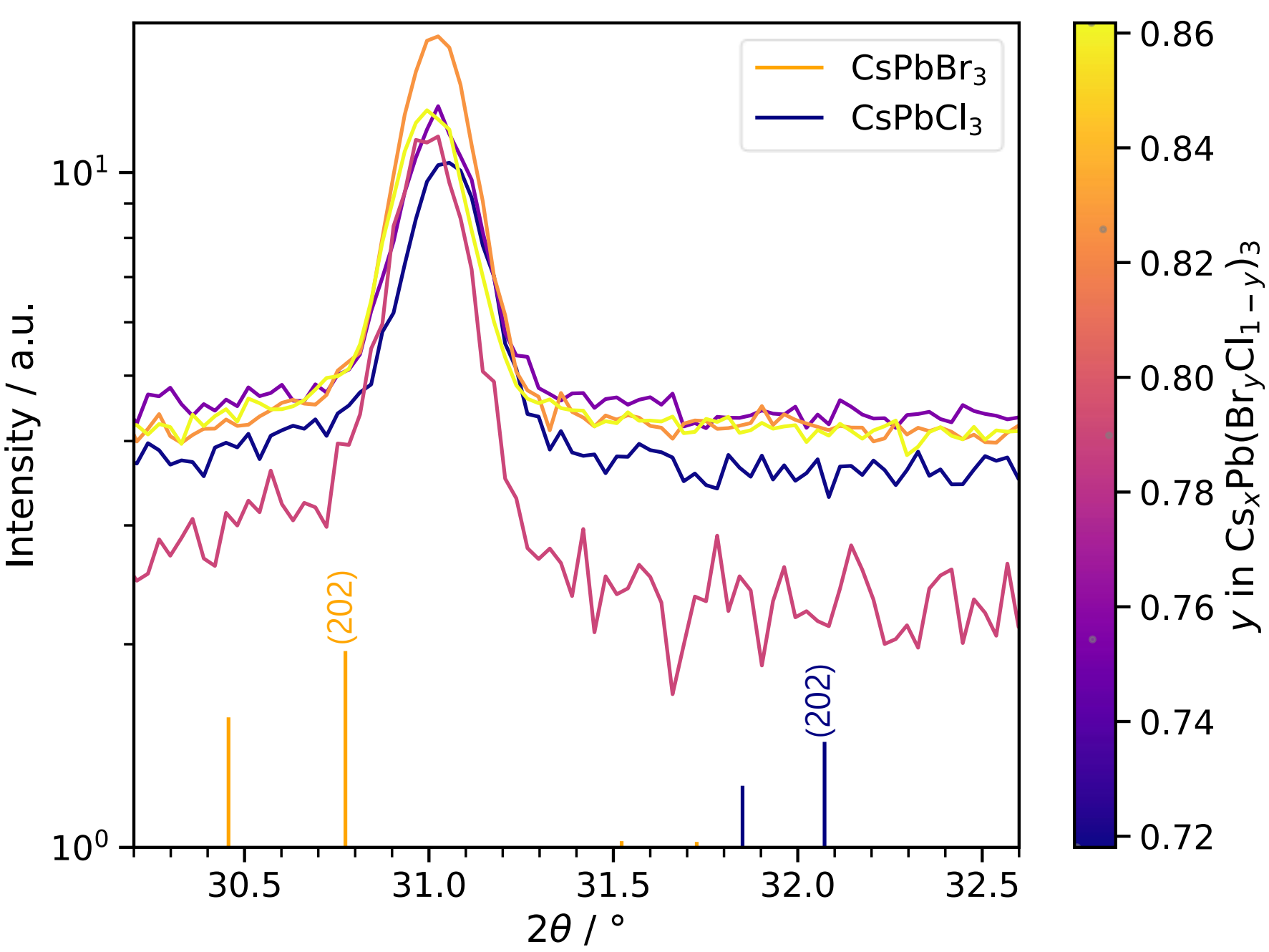

**Figure S4.** Detailed diffractogram in the 30.2° < $2\theta$ < 32.6° range for selected samples spanning equidistant variations in the halide ratio. The simulated powder XRD patterns based on references for $CsPbBr_3$ (ICSD-14608) and $CsPbCl_3$ (ICSD-143611) are specified as vertical lines.

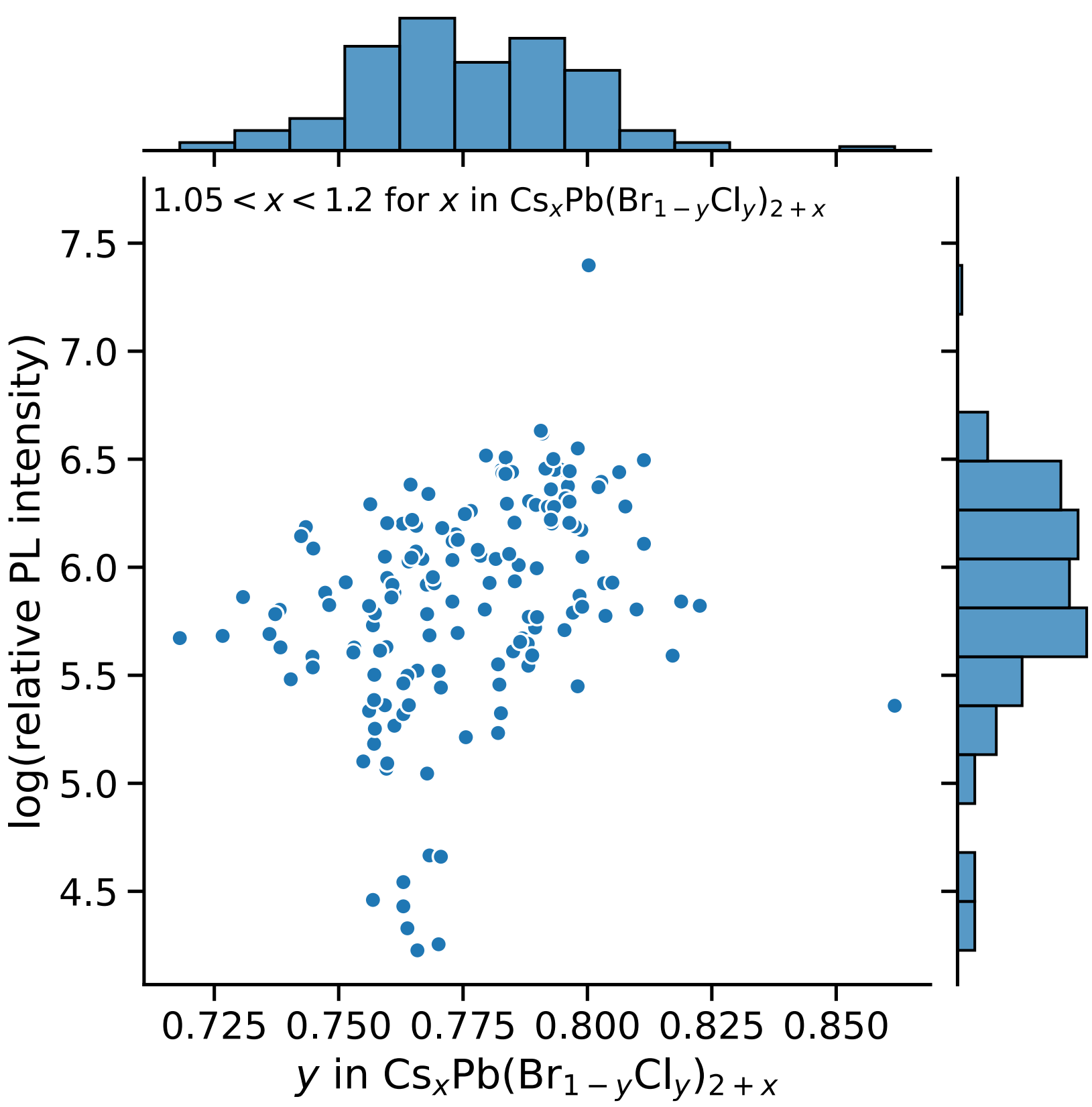

**Figure S5.** Changes of the relative PL intensity for different halide ratios (*i.e.*, $y$ in $Cs_xPb(Br_{1-y}Cl_y)_{2+x}$). The Cs/Pb ratios were confined to a range of 1.05 to 1.2.

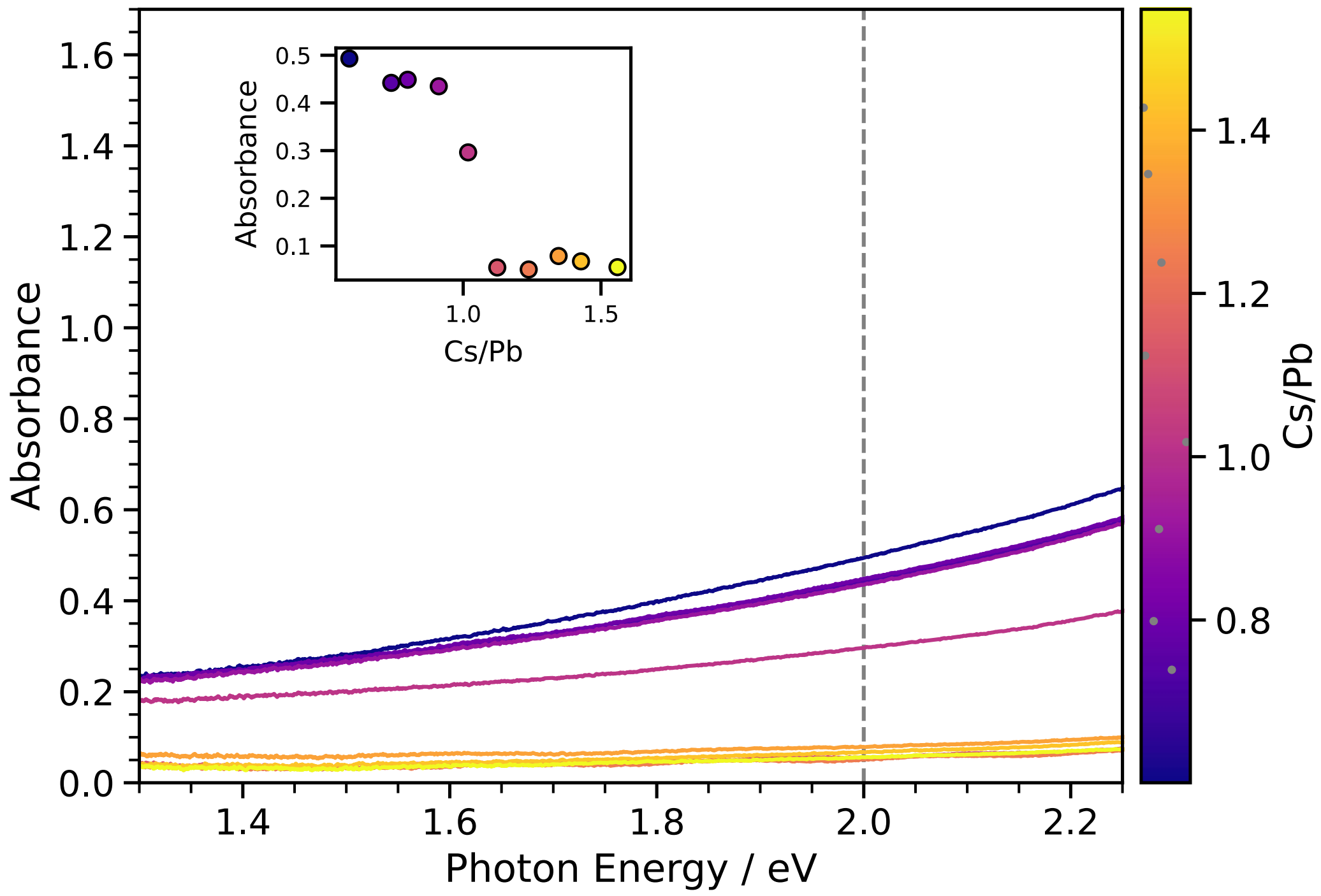

**Figure S6.** Absorbance of representative $Cs_xPbBr_{2+x}$ thin films with $x \geq 0.8$ at photon energies of 1.3 to 2.25 eV. The inset depicts the value at 2.0 eV with varying Cs/Pb ratios.

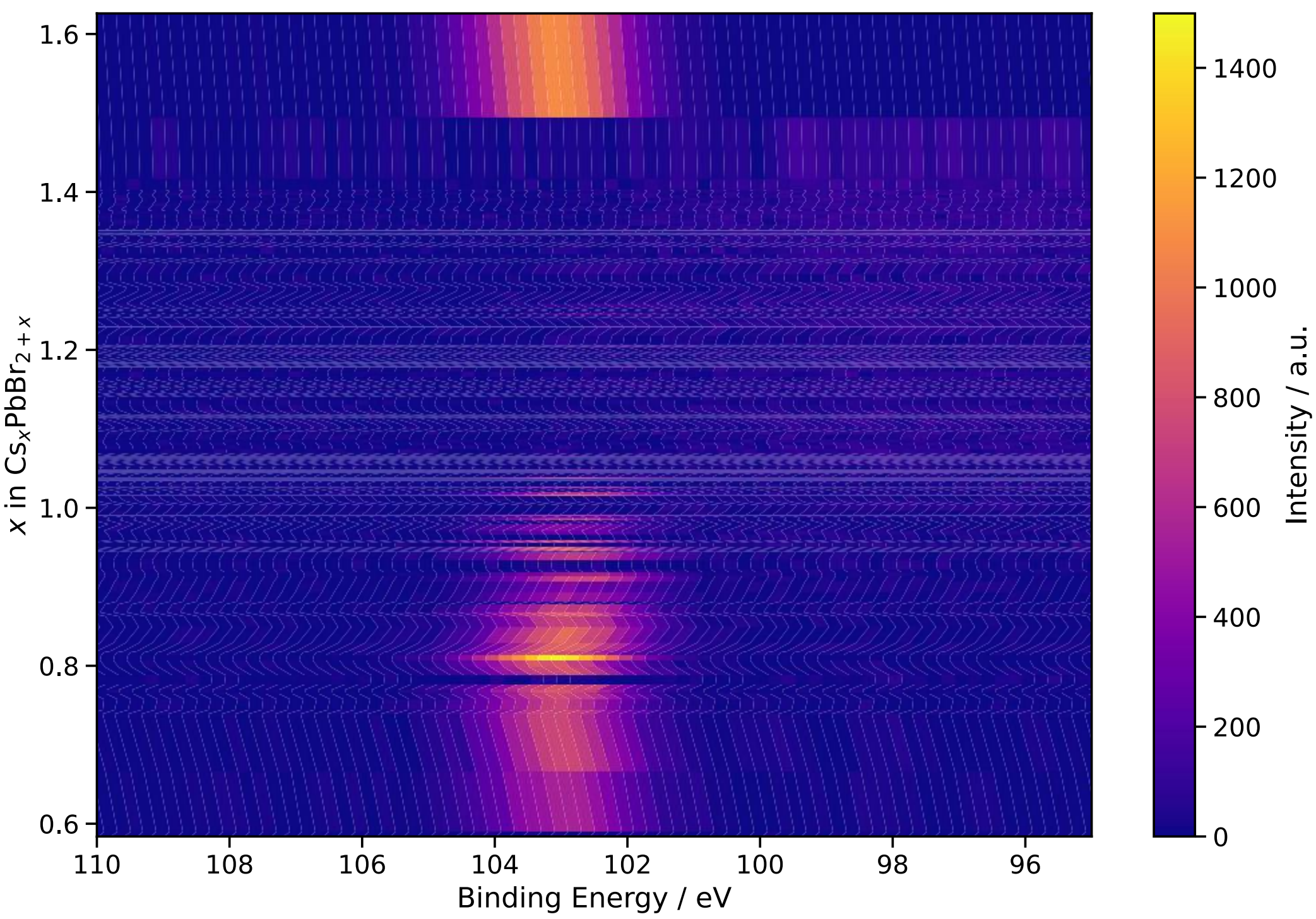

**Figure S7.** Si 2p core level spectra obtained from XPS across the $Cs_xPbBr_{2+x}$ compositional space. The Si signal originates from the substrate and indicates delamination.

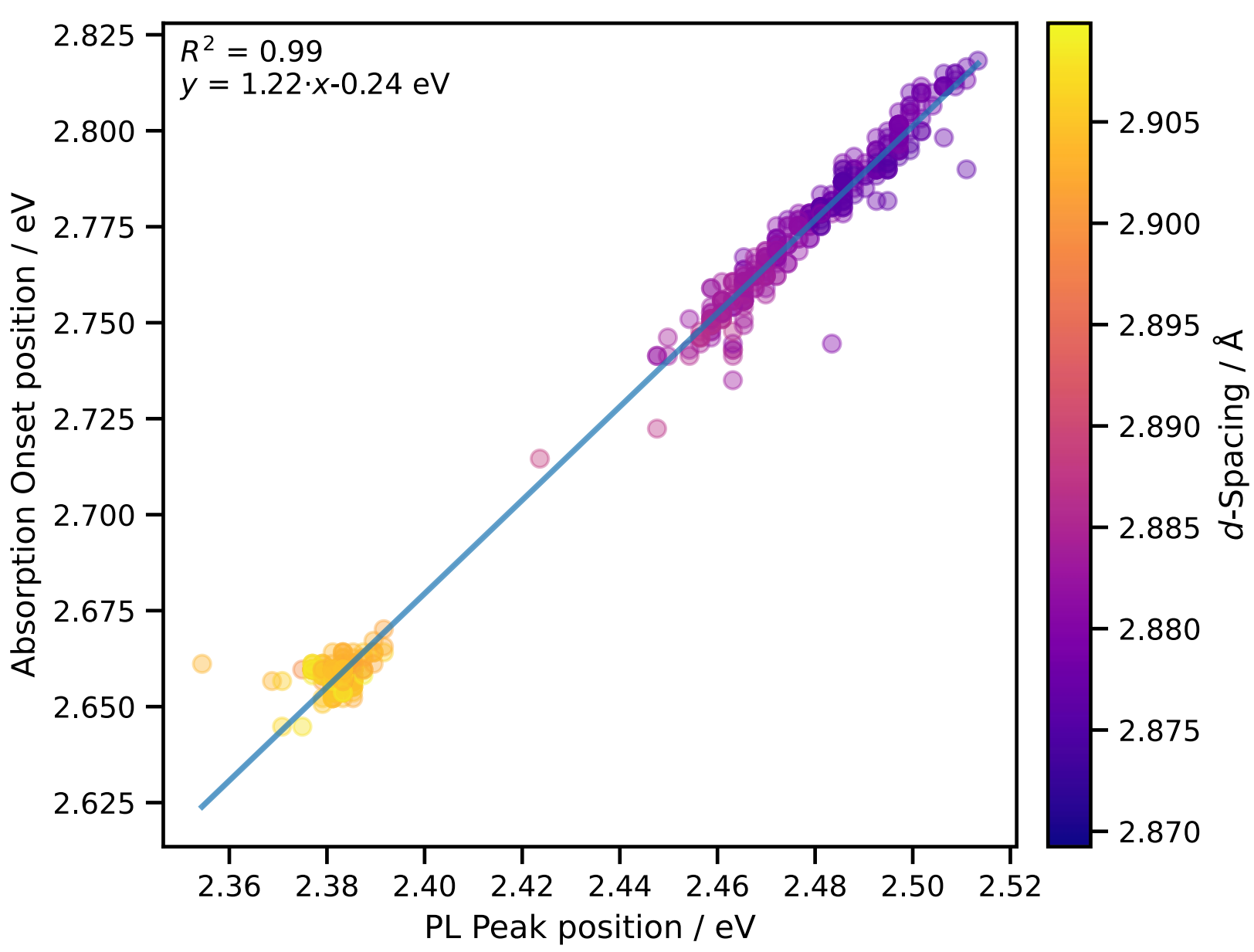

**Figure S8.** Linear correlation of the PL peak position *vs.* the Absorption Onset position as determined by the inflection point of the absorption. The inflection point method is robust for high-throughput workflows but leads to a systematic error in band edge determination. This explains the rather large apparent Stokes shift of over 0.2 eV. The slope of nearly 1 between the Absorption Onset position and the PL peak position suggests this apparent Stokes shift is similar across the explored compositional space.

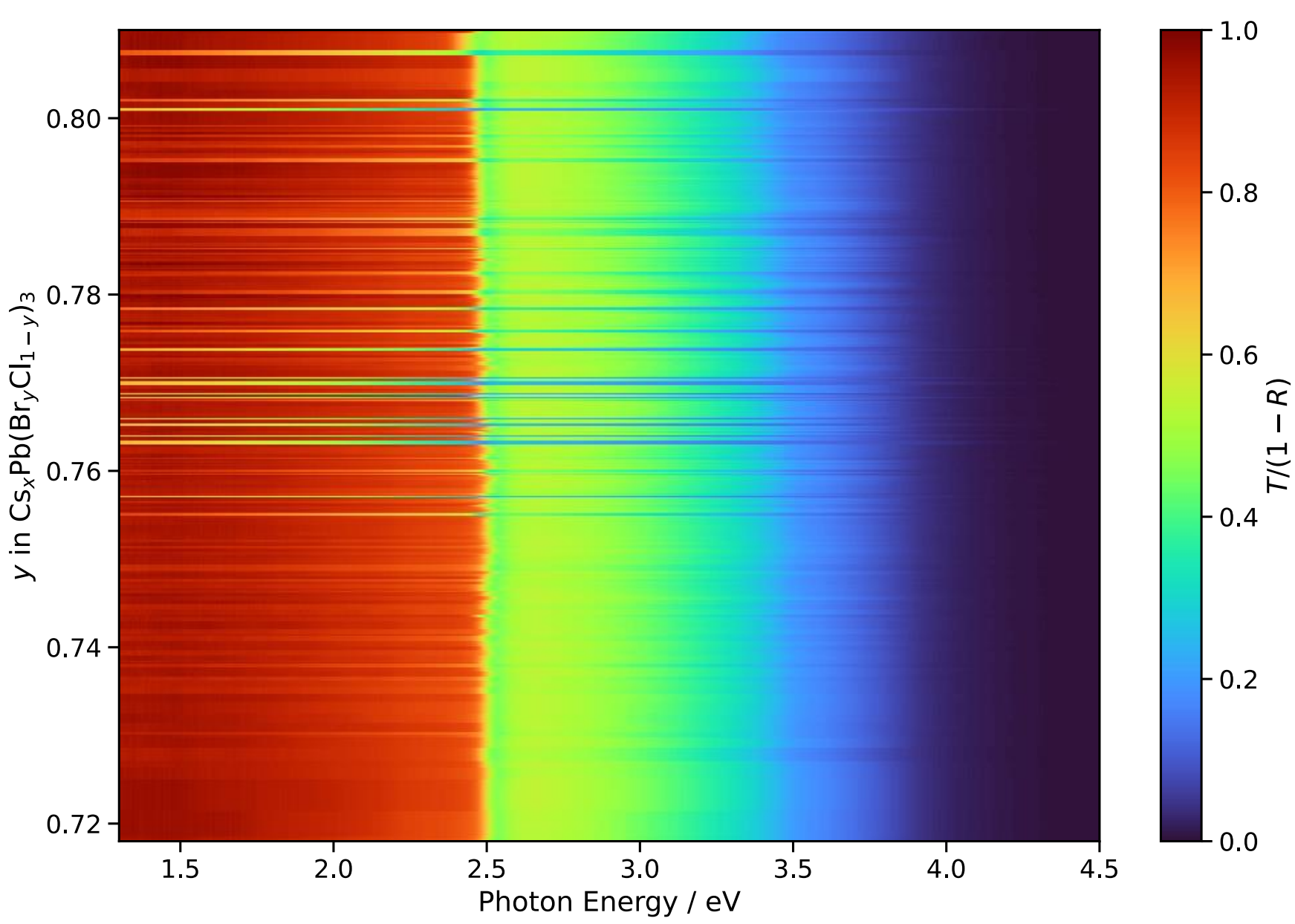

**Figure S9.** False color plot depicting changes in the reflection-corrected transmission ($T$ = Transmittance, $R$ = Reflectance) across all samples in the $Cs_xPb(Br_{1-y}Cl_y)_{2+x}$ compositional space.

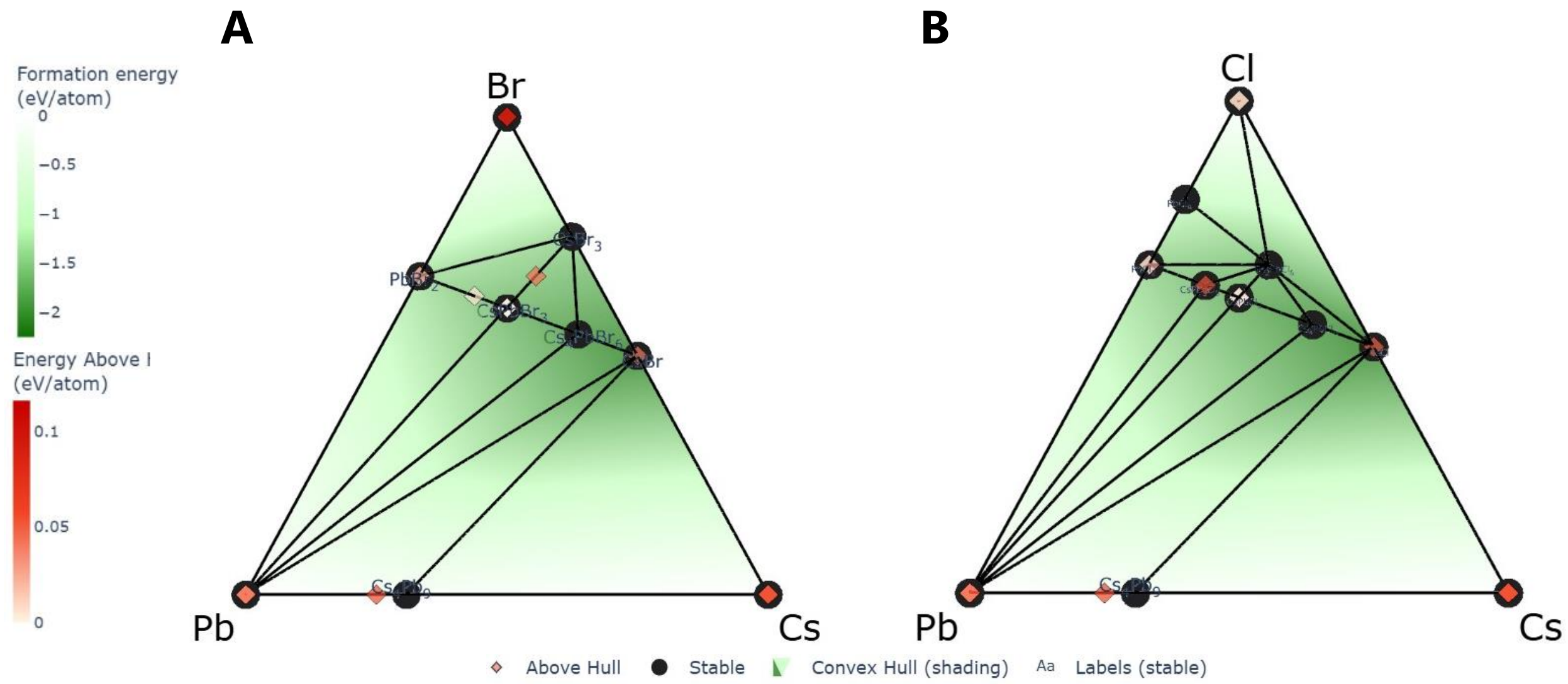

**Figure S10.** Phase diagram with formation energy as a contour plot for the **A)** Cs-Pb-Br, and **B)** Cs-Pb-Cl compositional space. Data extracted from Materials Project.[1]

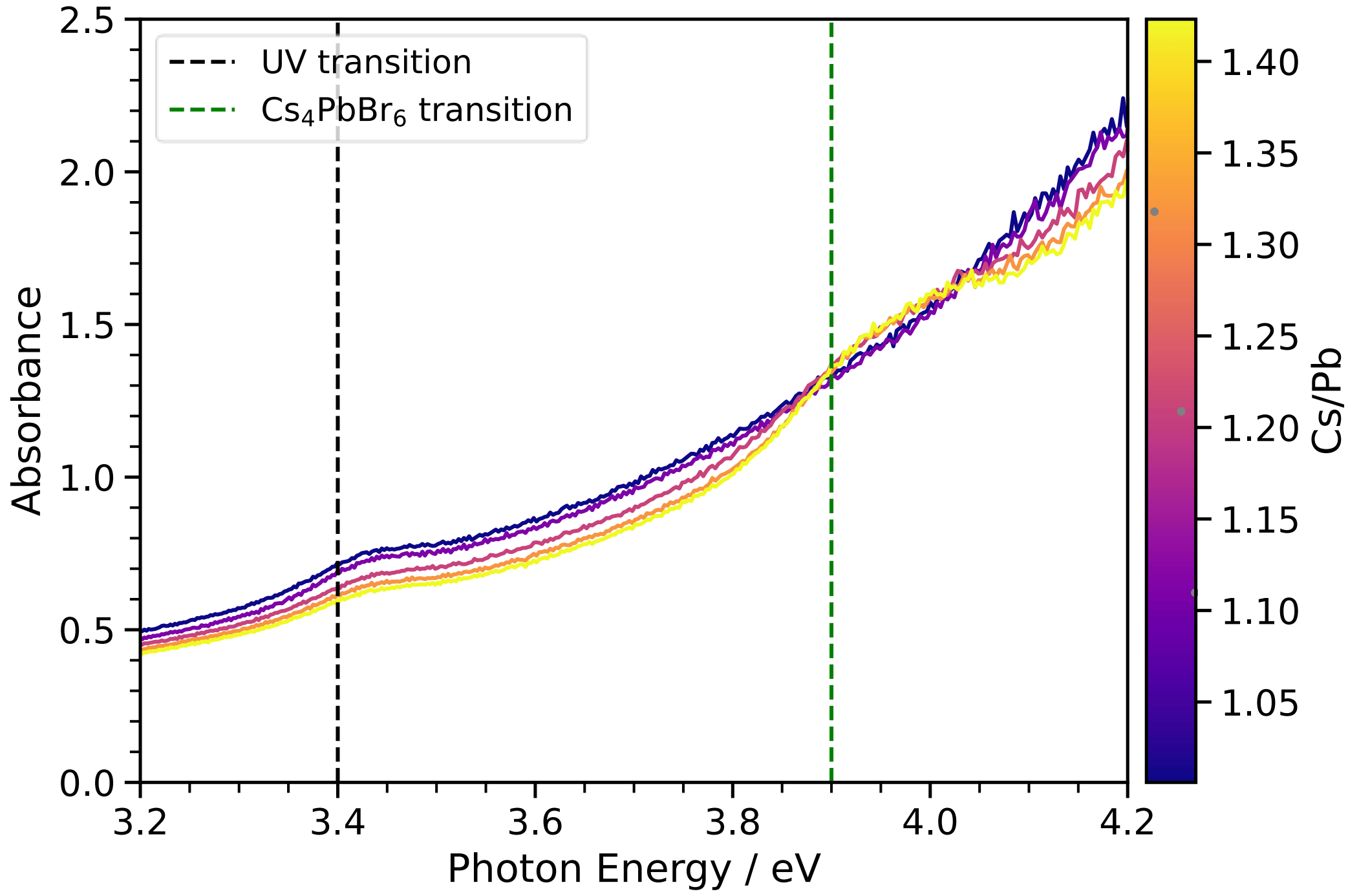

**Figure S11.** Absorbance of selected $Cs_xPbBr_{2+x}$ thin films with $x \leq 1$ at photon energies of 3.2 to 4.2 eV. The optical transition at 3.4 eV, referred to as UV transition in this report is marked in black, while the optical transition characteristic for $Cs_4PbBr_6$ is marked in green.

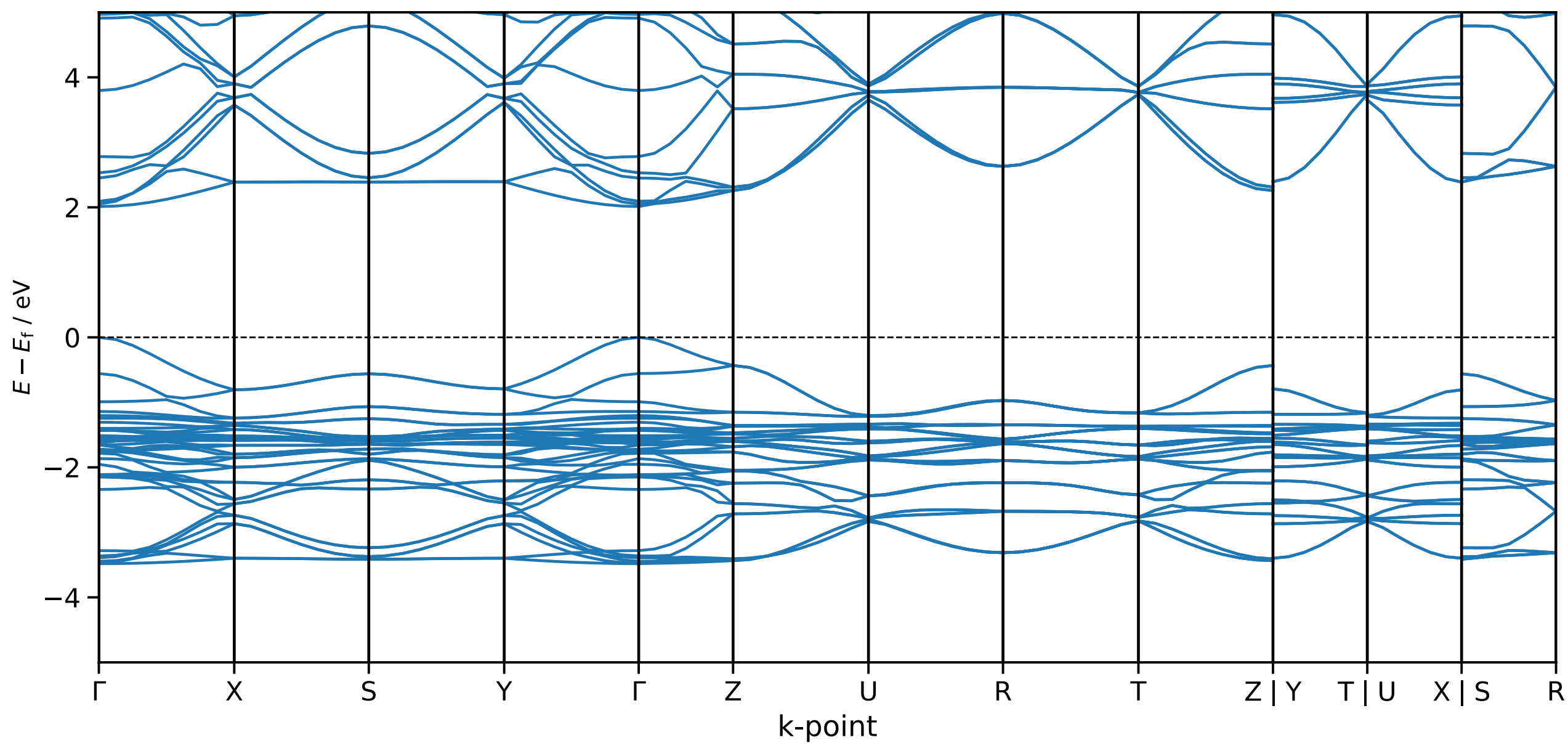

**Figure S12.** Band-dispersion diagram of orthorombic $CsPbBr_3$ as reported by Materials Project (mp-567629).[1,2]

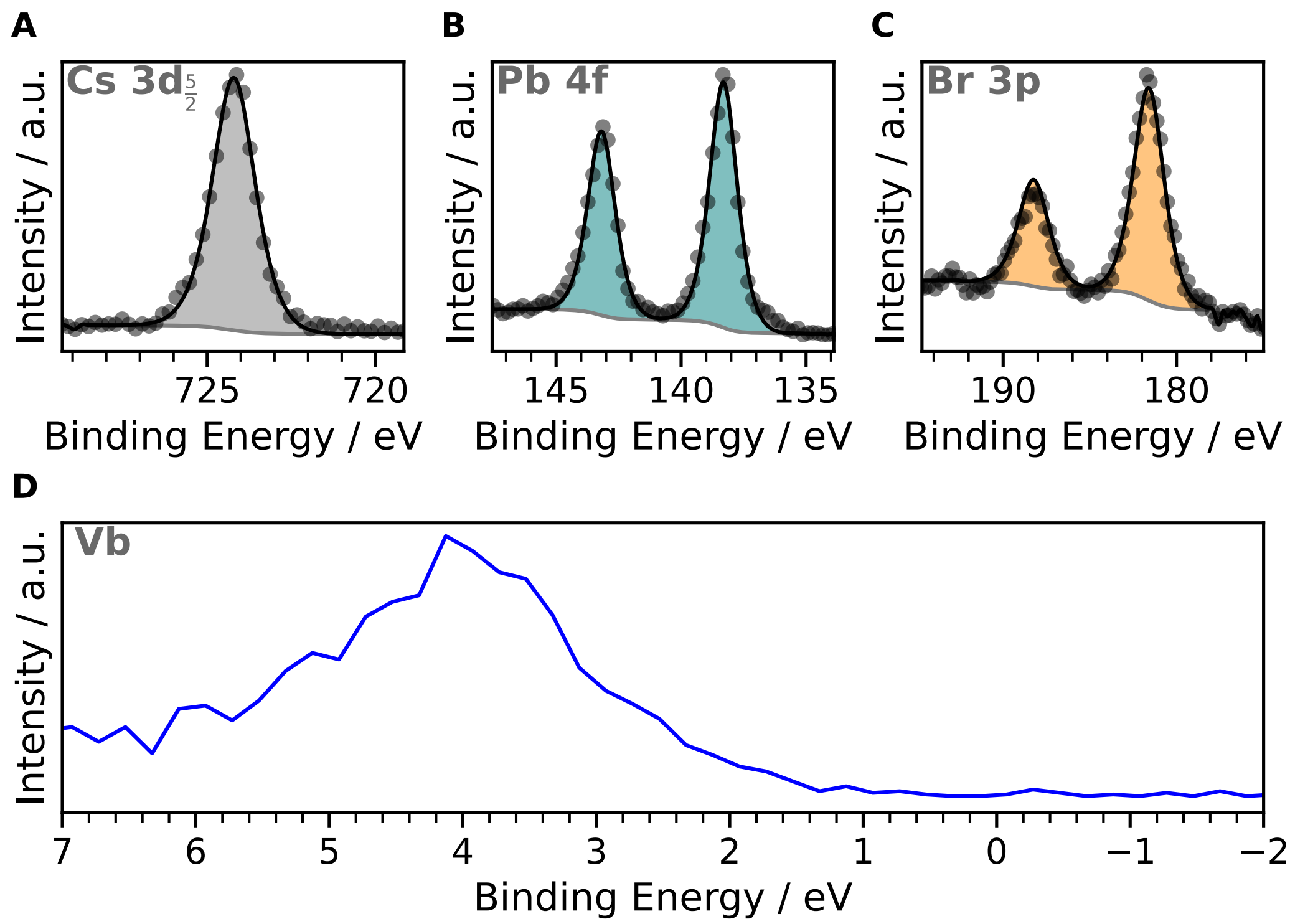

**Figure S13.** XPS results of a representative understochiometric ($x$<0.8) $Cs_xPbBr_{2+x}$ sample for the **A)** Pb 4f, **B)** Cs 3d, **C)** Br 3p, and **D)** valence band (Vb) regions.

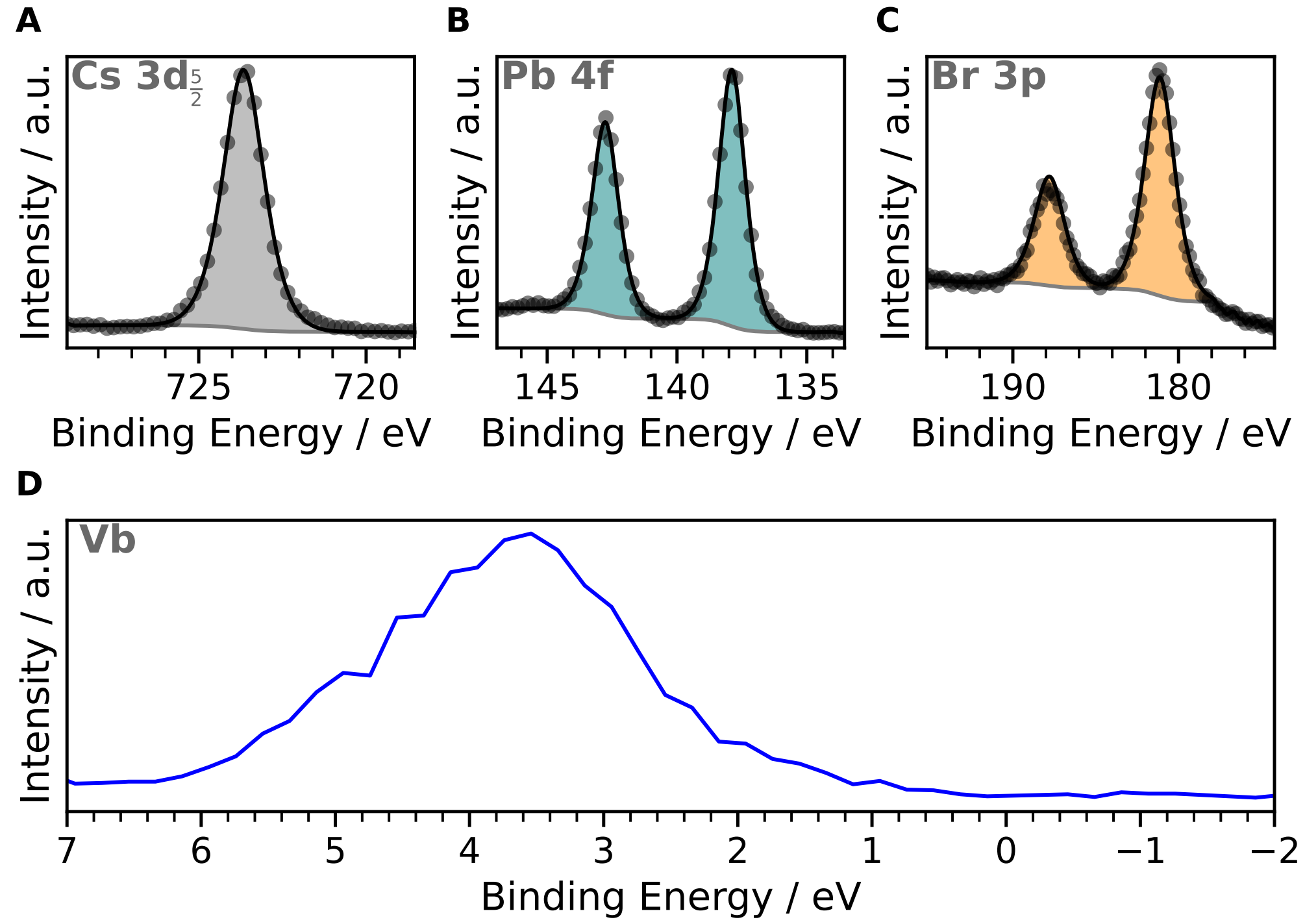

**Figure S14.** XPS results of a representative overstochiometric ($x$>1.2) $Cs_xPbBr_{2+x}$ sample for the **A)** Pb 4f, **B)** Cs 3d, **C)** Br 3p, and **D)** valence band (Vb) regions.

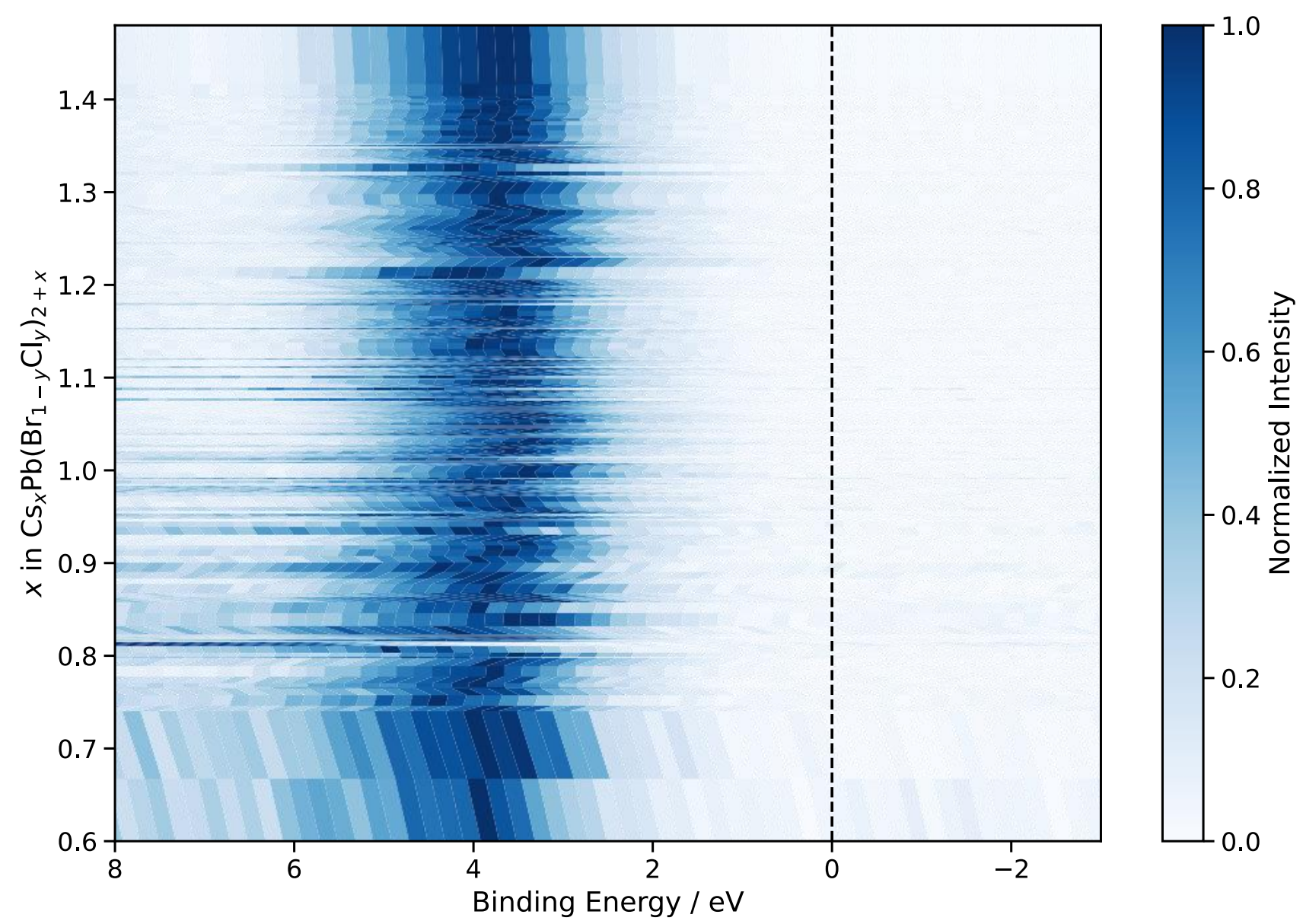

**Figure S15.** False-color plots of the valence band regions for all measured samples. The absence of features near the Fermi level (Binding Energy = 0 eV) is consistent with the lack of metallic phases in all samples.

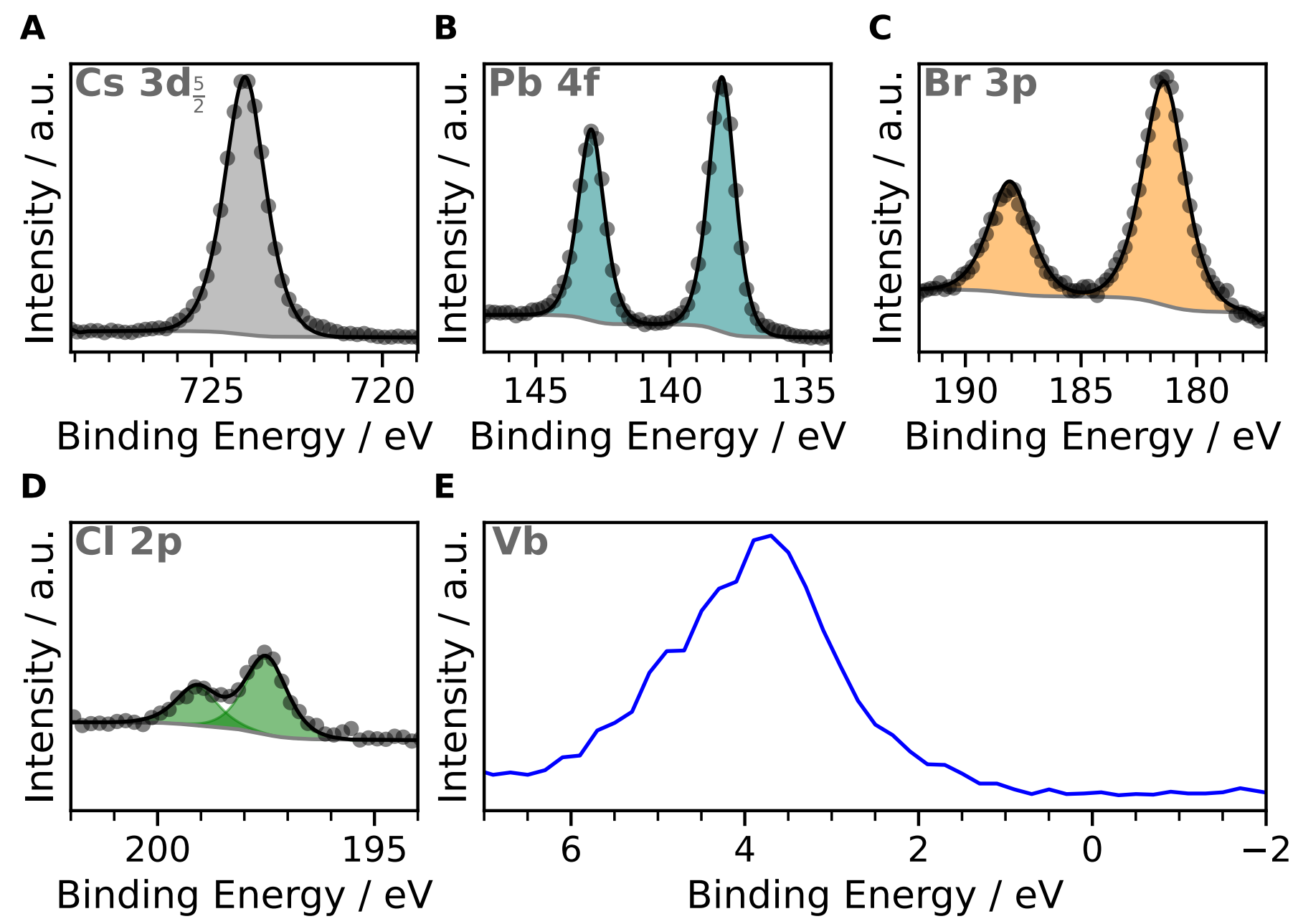

**Figure S16.** Detailed and fitted XPS spectra of the **A)** Cs 3d, **B)** Pb 4f, **C)** Br 3d, and **D)** Cl 2p core-level regions, as well as the **E)** valence band (Vb) region for a sample in the $Cs_xPb(Br_{1-y}Cl_y)_{2+x}$ compositional space.

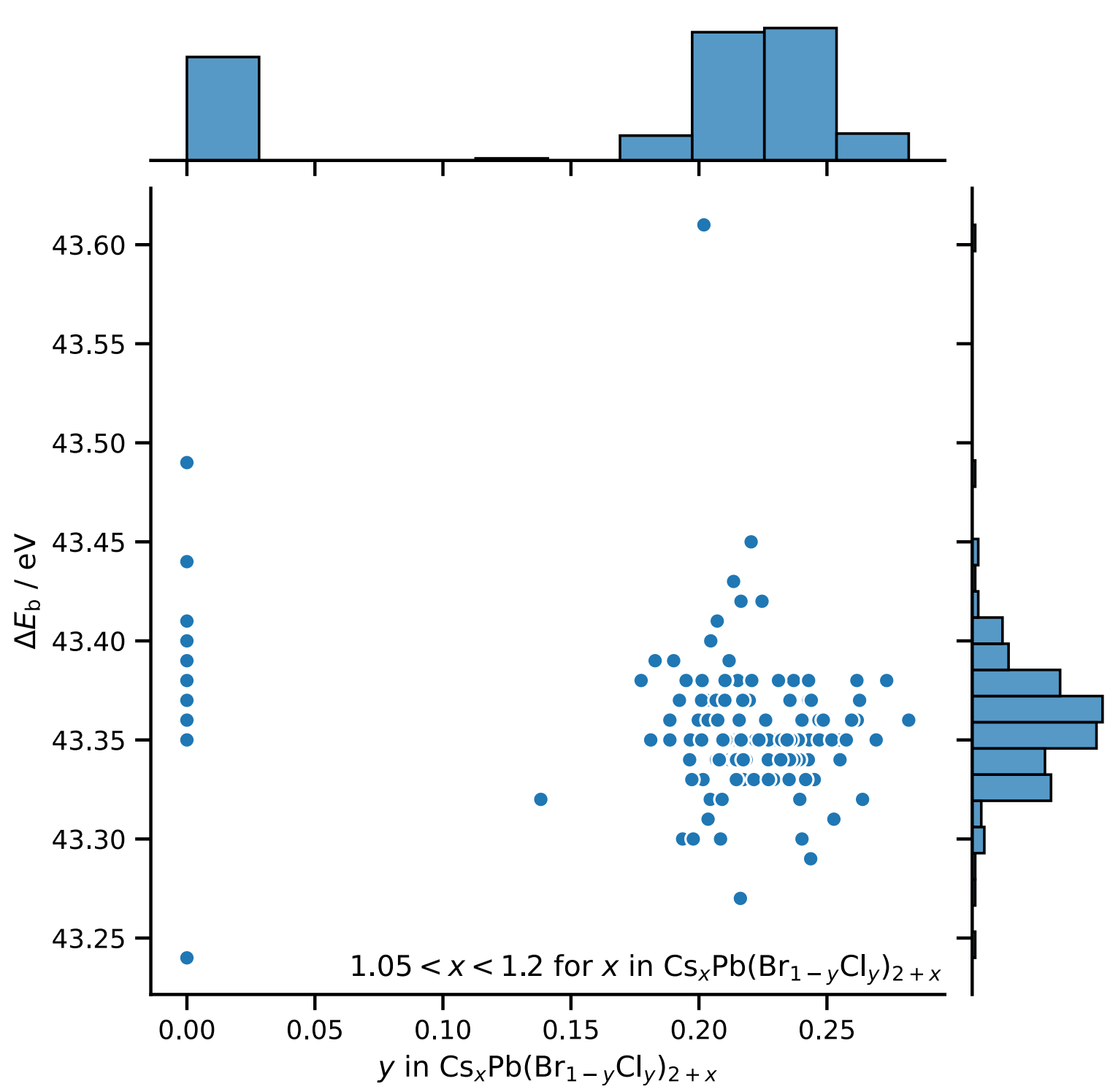

**Figure S17.** Changes of the relative binding energy $\Delta E_b = E_b(\text{Br } 3p_{3/2}) - E_b(\text{Pb } 4f_{7/2})$ for varying halide ratios (*i.e.*, $y$ in $Cs_xPb(Br_{1-y}Cl_y)_{2+x}$). The Cs/Pb ratios were confined to a range of 1.05 to 1.2.